\documentclass[reprint,nofootinbib,amsmath,amssymb,aps,prd]{revtex4-2}
\usepackage{graphicx}
\usepackage{dirtytalk} 
\usepackage[autostyle]{csquotes} 
\usepackage{geometry}
\usepackage[english]{babel}
\usepackage{amsmath}
\usepackage{physics}
\usepackage{tensor}
\usepackage{hyperref} 
\usepackage{cleveref}
\usepackage{amsfonts}
\usepackage{comment}
\usepackage{mathtools}

\usepackage{soul} 
\setstcolor{cyan} 

\usepackage{dcolumn} 
\usepackage{bm}      

\usepackage{xcolor}

\usepackage{orcidlink}

 \newcommand\pd[2]{\frac{\partial#1}{\partial#2}}
 \newcommand\lrb{\left(}
 \newcommand\rrb{\right)}

\newcommand\T{\rule{0pt}{2.6ex}}       

\usepackage{phfcc}

\begin{document}
\preprint{APS/123-QED}
\title{Asymptotically Schwarzschild solutions in \texorpdfstring{\boldsymbol{$f(R)$}}{f(R)} extension of general relativity}
\author{Federico Scali\orcidlink{0009-0004-0637-561X}}
\email{fscali@uninsubria.it}
\affiliation{Department of Science and High Technology, University of Insubria, Via Valleggio 11, 22100, Como, Italy}
\affiliation{INFN section Milan, Via Celoria 16, 20133 Milan, Italy}
\author{Oliver F. Piattella\orcidlink{0000-0003-4558-0574}}
\email{of.piattella@uninsubria.it}
\affiliation{Department of Science and High Technology, University of Insubria, Via Valleggio 11, 22100, Como, Italy}
\affiliation{INFN section Milan, Via Celoria 16, 20133 Milan, Italy}

\begin{abstract}
We address the question of how to build a class of $f(R)$ extensions of general relativity which are compatible with solar system experiments, without making any preliminary assumption on the properties of $f$.
The aim is reached by perturbatively solving the modified Einstein equations around a Schwarzschild background and retrieving \textit{a posteriori} the corresponding $f(R)$. This turns out to be nonanalytical in $R=0$ and should be intended as the leading correction to the Einstein-Hilbert action in the low curvature limit. The parameters characterizing the $f(R)$ class are then set by constraining the corrections to four different local tests with the observations.

The result is a class of $f(R)$ theories built up from a purely bottom-up approach and compatible with the local tests. At a more general level, this result can help constraining exact $f(R)$ models working in cosmology, since it provides the correct local limit. Further developments and possible extensions of the approach to cosmology are also discussed.
\end{abstract}

\maketitle

\tableofcontents

\section{Introduction}
It is curious and quite disturbing that the dark sector of the currently accepted cosmological model (the $\Lambda$-CDM model)  be the key to explain the dynamics of the universe and of galaxies and, at the same time, one of the greatest mysteries of contemporary physics. The dark sector encompasses two components of the energy density in the universe: \textit{cold dark matter} (DM) and \textit{dark energy} (DE). \newline
\indent DM is a pressureless perfect fluid constituting almost $85\%$ of the matter content in the universe \cite{Weinberg2008}. The first piece of evidence of DM effects came in $1933$  with the work of Zwicky \textit{et al}.
\cite{Zwicky1933}, who realized a large discrepancy between the mass to light ratio of the virialized COMA cluster and the mass to light ratios of the individual, visible galaxies within the cluster.
After four decades, the spectroscopic observations of the Andromeda Galaxy performed by Rubin and Ford  \cite{Rubin1970, Rubin1978} definitely opened the possibility that halos of nonluminous matter could surround the disk galaxies. The observations showed a profile of the rotation velocities extending flat well beyond the visible edge of the disk, thus enforcing the idea of a DM halo. Since then, DM has been invoked to explain a plethora of different phenomena, from the weak lensing of the Bullet Cluster \cite{Clowe2006} to the cosmological structure formation \cite{Weinberg2008}. See \cite{Piattella2018} for an overview on the main probes and \cite{Bertone2018} for an historical introduction. The main theoretical proposals as DM constituents include extensions of the Standard Model of particle physics \cite{Profumo2017} and primordial black holes \cite{Villanueva2021}. See also \cite{Cebri2023} for a review on the detection methods.\newline
\indent On the other hand, the conceptual roots of DE can be traced back to soon after the birth of general relativity (GR) when in 1917 \say{Cosmological Considerations on the General Theory of Relativity} \cite{Lorentz1952} (pp. 177$-$188) Einstein introduced the \textit{cosmological constant} (CC) $\Lambda$ in its field equations. At the cosmic scales a positive CC has the same effect as a perfect fluid with negative pressure, allowing a static universe with nonzero energy density to be a solution of the Einstein equations (EE). 
Although the possibility of a static universe was definitely discarded by the  observations of Hubble \cite{Hubble1929}, showing that the universe is actually expanding, the CC contribution continued to \enquote{rear its ugly head}\footnote{This citation is taken from Gamow's autobiography \cite{Gamow1970}.} for decades. This was because of the (possibly gravitating) vacuum fluctuations of the quantum fields, which take exactly the CC form given a Lorentz invariant vacuum\footnote{The first intuition in this direction came in 1968 from Zel'dovich, whose primary focus was to disprove the necessity of a vanishing zero point energy of the quantum fields and thus paved the way to a new field of activity in theoretical physics. In his words \enquote{The genie has been let out of the bottle, and it is no
longer easy to force it back in} \cite{Zeldovich1968}.} \cite{Zeldovich1968}. Meanwhile, the development 
of inflationary models in the 1980s \cite{Guth1981} also strengthened the importance of fluids with negative pressure. But the true revolution only happened toward the end of the century. Between $1998$ and $1999$, the members of the Supernova Cosmology Project \cite{Perlmutter_1999} and the High-Z Supernova Search Team \cite{Riess_1998} analyzed the luminosity-redshift relation of a number of Type Ia supernovae and discovered that the expansion of the universe is  accelerating.\footnote{From that moment onward, the cosmic fluid responsible for the acceleration was dubbed dark energy; see \cite{Calder2008} for a historical introduction.} Their best fit within the $\Lambda$-CDM model set the contribution of the CC to $|\rho_\Lambda| \simeq 10^{-47}\text{GeV}^4$, amounting to about $70\%$ of the total energy density in the universe.
This, of course, boosted renewed interest in the long-standing problem of a satisfactory theoretical explanation for the value of the CC but could not, and to date it still cannot, be settled within the particle physics framework. The reason is that, on the one hand, the \enquote{bare} CC appearing in the EE should cancel the regularized vacuum energy contribution from the matter fields with an insane degree of accuracy\footnote{This seems to happen whatever regularization procedure is employed \cite{Martin_2012}.} \cite{Weinberg1989} (\textit{fine-tuning problem}), and on the other hand, the phase transitions of the fundamental interactions which supposedly took place in the early universe dramatically change the classical contributions from the minima of the interaction potentials \cite{Martin_2012} (\textit{classical problem of phase transitions}). Because of these problems, alternative DE models which do not foresee the CC have been developed in the past $20$ years; see \cite{Tsujikawa2010, Bamba2012} for a complete compendium. Also, in July 2023 the European Space Agency has successfully launched the Euclid satellite with the primary aim of detecting new signatures of DE \cite{Amendola2018}.

Understanding the nature of DM and of DE is no doubt the greatest open problem in cosmology.
Evidently, the conceptual element shared by DM and DE is a certain use of GR to describe their gravitational interaction: in the Newtonian limit in the case of the galaxies and constrained by the cosmological principle in the case of the whole universe. One could argue that such restrictions on the use of GR are simply too strong and that, in fact, Einstein's theory could have some more to say if a more exact use were made \cite{Balasin2008,Crosta2020,Beordo2024,Lapi2023}. Another possibility which is worthy of consideration is that Einstein's theory may just fail to fully reproduce the dynamics of the gravitational field at those scales. If this were true, then a proper extension of GR could solve or, at least, alleviate the problem. We adopt this point of view from now on.

After the birth of GR it was soon clear that the geometrical approach to the description of the gravitational interaction could lead in principle to theories different and even more general than GR. Whether geometry is an essential feature of gravity was a matter of debate and remains an important, conceptual issue today, even if much less discussed. On this point we should mention that, although Einstein himself in $1919$ considered the possibility of modifying the theory in order to get trace free field equations\footnote{Interestingly, Einstein's primary focus was to stabilize the atomic structure of the electrons inside the atoms.} \cite{Lorentz1952} (pp.\ 189$-$198) (see also \cite{Fabris2023} for a brief review), he remained critical about the possibility of \enquote{geometrizing gravity} in an ontological sense \cite{Lehmkuhl2014}. The principle of equivalence, a physical principle, and its unification with the gravitational interaction being, in fact, the chief motors of its research \cite{Lehmkuhl2014}.

This being said, a remarkable step in the direction of a geometric extension of GR occurred in 1961 when, in the attempt to incorporate Mach's principle \cite{Mach1872}, Brans and Dicke introduced a scalar field as an additional mediator of the gravitational interaction, thus giving birth to the first \textit{scalar-tensor theory}\cite{Brans1961}. Thirteen years later, Hordenski in his doctoral thesis built the most general second-order scalar-tensor theory in four dimensions \cite{Horndeski1974}; see also \cite{Horndeski2024}. At that time 't Hooft and Veltman were proving the renormalizability of GR at one loop, finding corrections quadratic in the Ricci tensor and the scalar curvature to the Einstein-Hilbert action \cite{tHooft1974}. 
The same type of curvature corrections were encountered three years later by Davies \textit{et al.}\ who computed the one loop renormalized stress-energy tensor of a scalar field in a flat Friedmann universe \cite{Davies1977}. Using this result, in 1980 Starobinsky realized that a de Sitter era of accelerated expansion (or inflation) could solve the effective EE in the primordial universe \cite{Starobinsky1980}, when the one loop quantum corrections to the matter fields become relevant but the curvature is still too small to consider quantum effects of pure gravity. Therefore, when toward the end of the century the discovery of the accelerated expansion took place, a certain interest developed for higher derivative modified gravity models, with the perspective that nonlinear terms in the curvature could hide a natural explanation for the late times inflation. A minimal but quite general approach in this direction is to correct the usual Einstein-Hilbert action by an unspecified, nonlinear function $f$ of the scalar curvature $R$.\footnote{The first appearance is due to Buchdal \cite{Buchdahl1970}, who was indeed motivated by the shortcomings of the big bang model and the quantum-gravitational corrections to the Einstein-Hilbert action.} This is the so-called $f(R)$ \textit{gravity} and will be the theoretical framework of this work.

Complete and up to date accounts on $f(R)$ gravity can be found in \cite{Capozziello2010,Amendola_Tsujikawa2010,DeFelice2010,Nojiri2011,Nojiri2017} and references therein; see also \cite{Sotiriou2010} for a historical introduction to the main motivations. Hence, we will not linger on the technical aspects. Here we just want to emphasize that $f(R)$ gravity, as every extended theory of GR, must face two orders of problems. 
The first comes from the fact that GR is extremely well tested at the solar system scales \cite{Weinberg1972,Misner2017,Will2014}. This means that local tests as the gravitational redshift, the deflection of light by the Sun, the precession of closed orbits or the Shapiro delay can impose strong constraints on the possible class of $f$ functions and make it natural (even if not mandatory) to ask that $f(R)$ reduce to GR within the error bars at those scales. For this reason, conceptual tools such as the \textit{parametrized post-Newtonian} (PPN) expansion and the \textit{weak field} expansion are terribly helpful in the study of $f(R)$ gravity \cite{Capozziello2010,Capozziello2007,Capozziello2006,Capozziello2009}. As for the first, it should be mentioned that, under the crucial assumption that $f$ be analytical around the background value of $R$, two scenarios may happen \cite{Capozziello2010}. 
Either the effective range of the scalar degree of freedom introduced by $f(R)$ is larger than the solar system scales, in which case the theory predicts the value $\frac{1}{2}$ for the PPN parameter $\gamma$ and thus violates a severe experimental constraint \cite{Bertotti2003}, or the effective range is smaller than the solar system scales, in which case the $f(R)$ effects are hidden from the local tests and there is agreement with GR. Clearly, the problem in this case is that $f(R)$ would have no effects at the cosmological scales either, a reason why an adaptive range mechanism is usually introduced (the \textit{chameleon mechanism} \cite{Capozziello2010}).\newline 
\indent The second order of problems arises at the cosmological scales (we disregard the DM problem for the moment), since not only should the non linear terms in $f(R)$ naturally contribute to the late times inflation but also the theory should provide a radiation dominated epoch, followed by a matter dominated one \cite{Weinberg2008}.
The cosmological viability of $f(R)$ models has been extensively discussed in \cite{Carloni2005,Capozziello2006Cosmological,Amendola2007Conditions,Amendola2007Power,Amendola2007Are,Odintsov2017}. See also \cite{Olmo2005,Cognola2008,Nojiri2008} for discussions on the cosmological bounds on $f(R)$ gravity together with the solar system tests. \newline
\indent In this paper we address the first challenge of $f(R)$, that is, building a class of functions which is fully compatible with the local tests.
This is not a new task in the literature \cite{Capozziello2010}; however, general results are typically derived assuming $f(R)$ is analytical \cite{Jin2007, Capozziello2008} while particular models are studied starting off with a given form of $f$ as a function of $R$ \cite{Hu2007, Nojiri2007}. Here we take a different route: without making any preliminary assumption on the form of $f$, we study the modified EE outside a spherical source; we perturbatively solve them by asking for (i) full agreement with GR in the weak field limit and (ii) minimal regularity of the potentials and the derivative of $f$ as functions of the coordinates; and we retrieve \textit{a posteriori} the corresponding form of $f(R)$. \newline
\indent
More specifically, in Sec.\ \ref{sec: Einstein equations in f(R) gravity} we derive the modified EE for a general $f(R)$ in the metric formalism. In Sec.\ \ref{sec: Spherically symmetric systems} we specialize the equations to a static, spherically symmetric line element. 
For compatibility with GR we impose that the potentials reduce to the Schwarzschild ones far from the source. Moreover, to carry on explicit computations we ask that the corrections to the potentials be expandable in a Laurent series around the origin of the Schwarzschild coordinates. With these conditions we are able to perturbatively solve the modified EE, find the leading correction to the Schwarzschild line element, and retrieve \textit{a posteriori} the $f(R)$. This turns out to be non analytical in $R = 0$ and should be intended as the leading correction to the Einstein-Hilbert action in the low curvature limit, in the surroundings of a spherical source. The resulting $f(R)$ depends on two parameters: a universal coupling $c_1$ and an integer number $n$, which essentially determines the order of the correction. After a brief discussion on the PPN parameters of the theory in Sec.\ \ref{sec:PPN parameters}, we devote Sec.\ \ref{sec:solar system tests} to show how the parameters $c_1$ and $n$ can be fixed by the local tests. We compute the leading corrections to the gravitational redshift of sunlight, to the bending of light from a distant star by the Sun, to the precession of a closed orbit and to the Shapiro delay. In particular, we use measurements of the sunlight gravitational redshift \cite{GonzalezHernndez2020} to infer numerical bounds on $c_1$ at varying $n$. In the final Sec.\ \ref{sec:cosmology} we briefly look at cosmology.
We argue that, although the $f(R)$ found in this paper cannot directly be applied in that context, the same point of view and methodology can be employed to approach the problem. The results of this work would then serve as a consistency condition when local scales are reached.

\section{Einstein equations in \texorpdfstring{$f(R)$}{f(R)} gravity}
\label{sec: Einstein equations in f(R) gravity}
To fix the notation in this section we derive the modified EE for $f(R)$ gravity. We consider the action $S = S_G + S_m$, where 
\begin{equation}
    S_G =  \frac{1}{2k}\int_\mathcal{M} d^4x\sqrt{-g} f(R)
    \label{eq: gravitational action of f(R)}
\end{equation}
is the gravitational action and $S_m$ is the action of the matter fields. In $S_G$, $k = 8\pi G$ in natural units $(\hbar = c = 1)$, the integration is extended over the spacetime manifold $\mathcal{M}$ and $g$ is the determinant of the metric tensor, whose signature is mostly plus.\footnote{For the metric, the curvature tensors and the EE we use the Misner-Thorne-Wheeler \enquote{+,+,+,+} convention \cite{Misner2017}.} The function $f(R)$ is an arbitrary (and possibly nonlinear) function of the scalar curvature $R$. In particular, we do not preliminarily require $f$ to be analytical anywhere. Notice that for $f(R) \equiv R$ the usual Einstein-Hilbert action is recovered.  

We work in the metric formalism; hence $S_G = S_G[g]$ and the field equations for gravity are obtained by varying $S$ with respect to the metric. Variation of the matter action gives the energy-momentum tensor of the matter fields
\begin{equation}
    \delta_g S_m = \frac{1}{2}\int d^4x \sqrt{-g}T^{\mu\nu} \delta g_{\mu\nu}.
\end{equation}
The variation of $S_G$ hides a subtlety and we review it here. We get
\begin{equation}
\begin{aligned}
    \delta_g S_{G} =& \frac{1}{2k}\int_\mathcal{M}d^4x\sqrt{-g}\delta g_{\mu\nu}\left(\frac{1}{2}g^{\mu\nu} f(R) - \phi R^{\mu\nu}\right) \\& - 
    \frac{1}{2k}\int_\mathcal{M} d^4x\sqrt{-g}\phi(g^{\rho\sigma}g^{\mu\nu} - g^{\rho\mu}g^{\sigma\nu})\nabla_\rho \nabla_\sigma \delta g_{\mu\nu},
\end{aligned}
\end{equation}
where $\nabla$ denotes covariant differentiation and we have defined the scalar field $\phi \equiv \pd{f(R)}{R}$.\footnote{The derivative of $f$ embodies the additional, effective scalar degree of freedom of $f(R)$ gravity. This can be seen for example considering the O'Hanlon action $S_G = \frac{1}{2k}\int_\mathcal{M} d^4x\sqrt{-g} (\Phi R - V(\Phi))$ \cite{O'Hanlon1972}, in which an additional scalar field mediates gravity. Eliminating $\Phi$ through the field equations the $f(R)$ action is recovered.}

The second line can be split in a bulk plus a boundary part
\begin{equation}
\begin{aligned}
    &- 
    \frac{1}{2k}\int_\mathcal{M} d^4x\sqrt{-g}\phi(g^{\rho\sigma}g^{\mu\nu} - g^{\rho\mu}g^{\sigma\nu})\nabla_\rho \nabla_\sigma \delta g_{\mu\nu} =\\
    &-\frac{1}{2k}\int_\mathcal{\partial M} d^3y\sqrt{h}\,\alpha\, n_\rho[\phi(g^{\rho\sigma}g^{\mu\nu} - g^{\rho\mu}g^{\sigma\nu}) \partial_\sigma \delta g_{\mu\nu}]
    \\ &+ 
    \frac{1}{2k}\int_\mathcal{M} d^4x\sqrt{-g}
    (\nabla_\rho\phi)(g^{\rho\sigma}g^{\mu\nu} - g^{\rho\mu}g^{\sigma\nu}) \nabla_\sigma \delta g_{\mu\nu},
\end{aligned}
\label{eq: splitting of the term coming from the variation of the Ricci tensor in bulk plus boundary parts}
\end{equation}
where $\{y_a\}$ are the proper coordinates of the boundary $\partial \mathcal{M}$, which we assume to be nowhere null, and $h_{\mu\nu}(x) = \pd{y^a}{x^\mu}\pd{y^b}{x^\nu} h_{ab}(y)$ is its induced metric,\footnote{See \cite{Poisson2009} for an introduction to the formalism of embedded surfaces.} which we held fixed during the variation. The vector $n$ is the unit normal vector field to the boundary and $\alpha$ is a number taking values $+1$ for outgoing $n$ and $-1$ for ingoing $n$. In the derivation we have also used that by hypothesis $\delta g|_{\partial\mathcal{M}} = 0$, and therefore $\nabla \delta g|_{\partial\mathcal{M}} = \partial \delta g |_{\partial\mathcal{M}}$. At the boundary we can write $g_{\mu\nu} = \epsilon n_\mu n_\nu + h_{\mu\nu}$, with $\epsilon = +1$ $(-1)$ if $\partial\mathcal{M}$ is timelike (spacelike); hence,
\begin{multline}
    -\frac{1}{2k}\int_\mathcal{\partial M} d^3y\sqrt{h}\,\alpha\, n_\rho[\phi(g^{\rho\sigma}g^{\mu\nu} - g^{\rho\mu}g^{\sigma\nu}) \partial_\sigma \delta g_{\mu\nu}] =\\= -\frac{1}{2k} \int_{\partial \mathcal{M}} d^3y \sqrt{h}\, \alpha\, \phi\, h^{\mu\nu}n^\sigma \partial_\sigma\delta g_{\mu\nu}.
\end{multline}
Although this is a boundary contribution it is in general nonzero because the stationary action principle still allows for a nontrivial variation of the configuration variables orthogonal to the boundary surface. In GR the issue is solved by adding to the gravitational action a term proportional to the extrinsic curvature \cite{Poisson2009}. In $f(R)$ gravity the presence of the scalar $\phi$ makes things fairly more complicated (see, e.\ g.\ , \cite{Deruelle2010}). Here we take a practical point of view and assume that such a boundary contribution can, in fact, be eliminated by adding a proper counterterm to the gravitational action. 

Integrating by parts once the second term in (\ref{eq: splitting of the term coming from the variation of the Ricci tensor in bulk plus boundary parts}) we finally obtain the variation of the gravitational action (up to boundary contributions)
\begin{equation}
\begin{aligned}
\delta_g S_{G} = \frac{1}{2k}\int_\mathcal{M}d^4x\sqrt{-g}\delta g_{\mu\nu}&\left(\frac{1}{2}g^{\mu\nu} f(R) - \phi R^{\mu\nu} \right. \\& \left. - g^{\mu\nu}\nabla^2\phi + \nabla^\mu\nabla^\nu\phi\vphantom{\frac{}{}}\right).
\end{aligned}
\end{equation}
The modified EE are therefore
\begin{equation}
    \phi R_{\mu\nu} - \frac{1}{2}g_{\mu\nu}f(R) + g_{\mu\nu}\nabla^2 \phi - \nabla_\mu\nabla_\nu \phi = T_{\mu\nu}.
    \label{eq: Einstein equations for f(R)}
\end{equation}
If $\phi \equiv 1$ the usual EE are recovered. Moreover, notice that a Ricci-flat vacuum solution in GR is not a solution for every $f(R)$, since $\phi$ should satisfy: 
\begin{multline}
     \phi R_{\mu\nu} 
     +\left(g_{\mu\nu} \nabla^2 R - \nabla_\mu \partial_\nu R\right)\pd{\phi}{R} \\ + \left[g_{\mu\nu}(\partial R)^2 - \partial_\mu R \partial_\nu R\right]\frac{\partial^2 \phi}{\partial R^2}= \frac{1}{2}g_{\mu\nu}f(R),
\end{multline}
in the limit of zero Ricci curvature. A strong condition to ensure GR solutions would be  $\lim\limits_{R \to 0} \phi < +\infty$ and $\lim\limits_{R \to 0} f(R) = 0$. A weaker condition would be 
$\lim\limits_{R_{\mu\nu},R \to 0} R_{\mu\nu}\phi,\, R\pd{\phi}{R},\, R^2\frac{\partial^2 \phi}{\partial R^2} = 0$ and again $\lim\limits_{R \to 0} f(R) = 0$.

\section{Spherically symmetric systems}
\label{sec: Spherically symmetric systems}
In this section we specialize the discussion to the gravitational field produced in vacuum by a spherically symmetric source. As shown in Appendix \ref{appendix: Spherically symmetric metric}, for such systems there exist spherical coordinates centered on the source in which the metric takes the form 
\begin{equation}
 ds^2 = -e^{\nu(r,t)}dt^2 + e^{\mu(r,t)}dr^2 + r^2 d\Omega^2,
 \label{eq: line element of a spherical system in proper coordinates}
\end{equation}
where $\nu(r,t)$ and $\mu(r,t)$ are two arbitrary functions of time and radius and $d\Omega^2 = d\theta^2 + \sin^2(\theta)d\varphi^2$ is the metric on the two-sphere.  

By direct computation one can find the full set of independent modified EE (\ref{eq: Einstein equations for f(R)}) for this metric. We report them here for the convenience of the reader\footnote{In order, these are the $tt+rr$, $tt+\theta\theta$, $tr$ and $\theta\theta$ components of (\ref{eq: Einstein equations for f(R)}). The subscripts indicate partial differentiation with respect to the coordinates 
$\left( g(r,t)_r=\frac{\partial g(r,t)}{\partial r}\right)$.}

\begin{widetext}
\begin{equation}
        \frac{1}{2}\left(\frac{2\phi}{r}\nu_r + \frac{2\phi}{r}\mu_r + \mu_r\phi_r + \nu_r\phi_r- 2\phi_{rr}\right)+
        \frac{1}{2}e^{\mu-\nu} \left(\vphantom{\frac{1}{2}}\mu_t\phi_t + \nu_t\phi_t- 2\phi_{tt}\right) = 0,
    \label{eq: proper complete set of modified Einstein equations in f(R) with spherical symmetry - 1}
\end{equation}
 
\begin{equation}
       \frac{\phi}{r^2} e^\mu 
       +\frac{1}{2}\left(\phi\nu_{rr} - \frac{\phi}{2}\nu_r\mu_r+\frac{\phi}{2}\nu_r^2 + \frac{\phi}{r}\nu_r + \frac{\phi}{r}\mu_r + \phi_r\nu_r - \frac{2\phi_r}{r} - \frac{2\phi}{r^2}\right) +
         \frac{1}{2}e^{\mu-\nu}\left(-\phi\mu_{tt} + \frac{\phi}{2}\mu_t\nu_t - \frac{\phi}{2}\mu_t^2 + \nu_t\phi_t - 2\phi_{tt}\right) = 0,
    \label{eq: proper complete set of modified Einstein equations in f(R) with spherical symmetry - 2}
\end{equation}

\begin{equation}
        \frac{\phi}{r}\mu_t - \phi_{tr}+ \frac{1}{2}\nu_r\phi_t + \frac{1}{2}\mu_t\phi_r = 0,
    \label{eq: proper complete set of modified Einstein equations in f(R) with spherical symmetry - 3}
\end{equation} 

\begin{equation}
        \frac{1}{2}e^\mu f(R) 
        + \frac{1}{2}\left(\phi\nu_{rr} - \frac{\phi}{2}\nu_r\mu_r + \frac{\phi}{2}\nu_r^2 + \frac{2\phi}{r}\nu_r +\mu_r\phi_r - \frac{4\phi_r}{r}-2\phi_{rr}\right)+
        \frac{1}{2}e^{\mu-\nu}\left(-\phi\mu_{tt}+ \frac{1}{2}\phi\mu_t\nu_t-\frac{\phi}{2}\mu_t^2+\mu_t\phi_t\right) =0.
    \label{eq: proper complete set of modified Einstein equations in f(R) with spherical symmetry - 4}
\end{equation}
\end{widetext}
As a warm-up we preliminarily study the equations in some simple cases.
\subsection{Affine GR}
We consider first the case
\begin{equation}
    \phi \equiv \text{const} \implies f(R) = \phi R + \Lambda.
    \label{eq: linear f(R)}
\end{equation}
From (\ref{eq: proper complete set of modified Einstein equations in f(R) with spherical symmetry - 3}) we get  $\mu(r,t) = \mu(r)$, while from (\ref{eq: proper complete set of modified Einstein equations in f(R) with spherical symmetry - 1})
\begin{equation}
    \nu(r,t) = -\mu(r) + T(t).
\end{equation}
Because of the freedom in the time reparametrization we can just consider
\begin{equation}
    \nu(r,t) = - \mu(r) \equiv \ln(N(r)),
\end{equation}
where a new function $N(r)$ has been defined for convenience. 
Using (\ref{eq: Ricci scalar for the metric of a spherical system}), (\ref{eq: linear f(R)}) and subtracting (\ref{eq: proper complete set of modified Einstein equations in f(R) with spherical symmetry - 4}) from (\ref{eq: proper complete set of modified Einstein equations in f(R) with spherical symmetry - 2})  we get a simple equation for $N$
\begin{equation}
    N_{rr} + \frac{2}{r}N_{r} = \frac{\Lambda}{\phi}.
\end{equation}
The solution to the homogeneous part is 
\begin{equation}
    N_0(r) = c_1 + \frac{c_2}{r^2},
\end{equation}
while a particular solution is given by
\begin{equation}
    N_p(r) = \frac{1}{6}+ \frac{\Lambda}{\phi}r^2.
\end{equation}
Therefore, the general solution for $\mu,\nu$ is 
\begin{equation}
    \nu(r) = \ln\left(c_1 + \frac{c_2}{r} + \frac{1}{6}\frac{\Lambda}{\phi}r^2\right) = -\mu(r)
\end{equation}
and the metric (\ref{eq: line element of a spherical system in proper coordinates}) takes the form 
\begin{equation}
    ds^2 = - \left(c_1 + \frac{c_2}{r} + \frac{1}{6}\frac{\Lambda}{\phi}r^2\right) dt^2  + \frac{1}{c_1 + \frac{c_2}{r} + \frac{1}{6}\frac{\Lambda}{\phi}r^2}dr^2 + r^2 d\Omega^2.
\end{equation}
The constants $c_1,c_2$ can be fixed asking that for $\Lambda = 0$, $g_{00}$ tend to $-1-2\phi_N$, where $\phi_N$ is the Newtonian potential of the source $\phi_N = -\frac{MG}{r}$. This implies $c_1 = 1$ and $c_2 = -2MG$; that is, the metric takes the form of the usual Schwarzschild$-$de Sitter solution with $\Lambda$ (related to) the cosmological constant.

\subsection{Schwarzschild solution}
Here we assume the Schwarzschild potentials 
\begin{equation}
    \nu = \ln\left(1-\frac{c}{r}\right) = -\mu,\,\,\,\, c\in \mathbb{R}
\end{equation}
and deduce which conditions this solution implies on $\phi|_{R = 0}$ and $f(0)$.

Equation (\ref{eq: proper complete set of modified Einstein equations in f(R) with spherical symmetry - 3}) gives
\begin{equation}
    \phi_{tr} - \frac{1}{r}\nu_r\phi_t = 0,
\end{equation}
which reduces to an identity if $\phi_t = 0$. However, let us assume $\phi_t \neq 0$. Integrating we get 
\begin{equation*}
    \nu(r) - \nu(r_0) = 2\ln(\phi_t) + g(t),
\end{equation*}
which by hypothesis would imply
\begin{equation*}
\phi = \sqrt{1-\frac{c}{r}} G(t) + h(r),
\end{equation*}
where $G(t) = \int_0^t dt' e^{-\frac{1}{2}g(t') - \frac{1}{2}\nu(r_0)}$. 
Inserting now Eq.\ (\ref{eq: proper complete set of modified Einstein equations in f(R) with spherical symmetry - 1})
and the expression for $\phi$ into Eq.\ (\ref{eq: proper complete set of modified Einstein equations in f(R) with spherical symmetry - 2}), after some algebra we get 
\begin{equation*}
    \left[\frac{c}{r^2}\sqrt{\frac{c}{r-c}}\frac{4c-3r}{2r(r-c)}\right]G(t) + \frac{3c-2r}{2r(r-c)}h_r + h_{rr} = 0,
\end{equation*}
which evidently cannot be satisfied at every time for every $r$. Let us take $\phi_t = 0$ then. Equation (\ref{eq: proper complete set of modified Einstein equations in f(R) with spherical symmetry - 1}) implies
\begin{equation}
\phi(r) = \alpha r + \beta, \quad \alpha,\beta \in \mathbb{R},
\end{equation}
while from (\ref{eq: proper complete set of modified Einstein equations in f(R) with spherical symmetry - 2})
\begin{equation}
\alpha \left[\frac{1}{2(r-c)} - \frac{3}{2}\frac{1}{r}\right] = 0.
\end{equation}
The latter is satisfied only if $\alpha = 0$, which implies $\phi = \beta$. Finally, Eq.\ (\ref{eq: proper complete set of modified Einstein equations in f(R) with spherical symmetry - 4}) implies $f(0) = 0$. This proves that the Schwarzschild metric is still a solution of the $f(R)$ extension if the strong condition stated at the end of Sec.\ \ref{sec: Einstein equations in f(R) gravity} is met.

\subsection{Static scalar field}
\label{subsec: Static scalar field}
Since it will be the main case of study, we show explicitly that a static scalar $\phi_t = 0$ implies a static metric. 

From Eq.\ (\ref{eq: proper complete set of modified Einstein equations in f(R) with spherical symmetry - 3}) 
\begin{equation}
    \mu_t\left(\frac{\phi}{r} + \frac{1}{2}\phi_r\right) = 0,
\end{equation}
and we see that either $\mu_t = 0$ or
$\phi = \phi_0\, \frac{r_0^2}{r^2}$.
However, the latter solution must be discarded because Eq.\ (\ref{eq: proper complete set of modified Einstein equations in f(R) with spherical symmetry - 1}) would imply $\phi_{rr} = 0$, which is absurd. Hence, we must take $\mu = \mu(r)$. In this case Eq.\ (\ref{eq: proper complete set of modified Einstein equations in f(R) with spherical symmetry - 1}) forces $\nu$ to take the form 
\begin{equation}
    \nu(r,t) = \nu_1(r) + \nu_2(t)
\end{equation}
but such a time dependence can always be reabsorbed into the redefinition of the temporal coordinate and we can simply take $\nu = \nu(r)$.

The discussion made here can be seen as a proof that the Birkhoff theorem \cite{Birkhoff1923} trivially holds in $f(R)$ gravity when $\phi$ is stationary but this is not true for more general $f(R)$ extensions \cite{Capozziello2010}.

\section{Asymptotically Schwarzschild solutions}
\label{sec:Asymptotically Schwarzschild solutions}
In this section we consider spherically symmetric, stationary systems with $\phi_t \equiv 0$. Since our focus is on the local tests, we refer to compact objects whose Schwarzschild radius sits well inside the visible radius $R_*$, as for the Earth or the Sun, so that we can always consider the surrounding gravitational field in the weak field limit. 
Einstein's theory is extremely successful in this setting; hence, we ask from the beginning a strong compatibility with GR where the gravitational field is weak. 

\subsection{Assumptions}

With this ideological posture it is natural to ask that asymptotically far from the source (in units of its Schwarzschild radius) the $f(R)$ extension be just a slight deviation from GR. As a consequence, we expect that the solution to the modified EE be in turn a slight deviation from the usual Schwarzschild metric in this region. We therefore look for solutions of the form 
\begin{equation}
    \begin{aligned}
        &\phi(r) = 1+\sigma(r),\\
        &\nu(r) = \ln\left(1-\frac{2MG}{r}\right) + g(r),\\
        &\mu(r) = - \ln\left(1-\frac{2MG}{r}\right) + m(r),
    \end{aligned}
    \label{eq: Schwarzschild potentials with arbitrary corrections}
\end{equation}
where $M$ is the mass of the compact object, equipped with the conditions
\begin{equation}
    \lim_{r/R_s\to\infty}\sigma(r),g(r),m(r) = 0, \,\,
    \lim_{r/R_s\to\infty} \frac{r}{R_s}g(r), \frac{r}{R_s}m(r) = 0,
    \label{eq: conditions on the functions of the metric in the quasi Schwarzschild solution}
\end{equation}
where $R_s = 2MG$ is the Schwarzschild radius. The conditions (\ref{eq: conditions on the functions of the metric in the quasi Schwarzschild solution}) just tell us that the additional functions fall off faster than the Schwarzschild potentials at infinity, possibly allowing for a perturbative derivation of the corrections to the Schwarzschild solution in the weak field region. 

As in the Schwarzschild case we take $\sigma, g, m$ analytical functions everywhere but in $r = 0$, where we assume a polar singularity. This means that they can be expanded around $r = 0$ in the Taylor-Laurent series 

\begin{equation}
    \begin{aligned}
    &\sigma(r) = \sum_{n=1}^{+\infty} \frac{\alpha^{\sigma}_n}{r^n} + 
    \sum_{m=0}^{+\infty} \beta^{\sigma}_m r^m,\\
    &g(r) = \sum_{i=2}^{+\infty} \frac{\alpha^{g}_i}{r^i} + 
    \sum_{j=0}^{+\infty}\beta^{g}_j r^j,\\
    &m(r) = \sum_{i=2}^{+\infty} \frac{\alpha^{m}_i}{r^i} + 
    \sum_{j=0}^{+\infty} \beta^{m}_j r^j,
    \end{aligned}
    \label{eq: Taylor-Laurent expansion of the corrections to the Schwarzschild potentials}
\end{equation}
with infinite convergence radius. Notice that the $\beta$ coefficients cannot be all positive or negative to allow condition (\ref{eq: conditions on the functions of the metric in the quasi Schwarzschild solution}). Moreover, if the Taylor part of the series is nonvanishing, the relevant asymptotic behavior of the potentials cannot be captured by means of a finite number of basis elements since they have an essential singularity at infinity. This really complicates any perturbative treatment of the solution, since it should be performed in terms of the arbitrary functions $\sigma,g,m$. 

Here we make the simplifying choice to restrict the solution to that class of functions with a vanishing Taylor part of the Laurent series, $\beta_i^{\sigma,g,m} = 0$ in (\ref{eq: Taylor-Laurent expansion of the corrections to the Schwarzschild potentials}). This allows a straightforward perturbative derivation of the leading correction to the Schwarzschild line element. Moreover, notice that this rules out Yukawa-like corrections, usually arising from the assumption of an analytical $f(R)$ \cite{Capozziello2010}.

\subsection{Leading correction to Schwarzschild}
Trailing the previous discussion, we consider the leading correction to the scalar 
\begin{equation}
    \phi(r) = 1 + \frac{c_1}{r^n} + \mathcal{O}\left(\frac{1}{r^{n+1}}\right).
    \label{eq: leading order correction to the derivative of f in the coordinates}
\end{equation}
Here $c_1$ should be intended as $c_1 = [c_1]\times l^n$, where $[c_1]$ is a real constant and $l$ is a fundamental length scale. We take $l$ so that condition (\ref{eq: conditions on the functions of the metric in the quasi Schwarzschild solution}) holds with good approximation right outside the radius of the star $R_*$; hence, at the very least, $c_1 \ll 2MG \times R_*^{n-1}$.

Equation (\ref{eq: proper complete set of modified Einstein equations in f(R) with spherical symmetry - 3}) is trivial in the stationary case. Equation (\ref{eq: proper complete set of modified Einstein equations in f(R) with spherical symmetry - 1}) provides
\begin{equation}
    \mu_r+\nu_r = \frac{2n(n+1)}{2-n}\left(\frac{1}{r} - \frac{r^{n-1}}{r^n + \frac{2-n}{2}c_1}\right) + \mathcal{O}\left(\frac{1}{r^{n+2}}\right),
\end{equation}
which can be integrated in $\mu+\nu$ giving
\begin{multline}
    \mu+\nu = \frac{2(n+1)}{n-2}\ln\left(1+\frac{2-n}{2}\frac{c_1}{r^n}\right) + \mathcal{O}\left(\frac{1}{r^{n+1}}\right) \\= -(n+1)\frac{c_1}{r^n} + \mathcal{O}\left(\frac{1}{r^{n+1}}\right).
    \label{eq: formula for mu + nu to order 1/r^n}
\end{multline}
Keeping into account the Laurent series (\ref{eq: Taylor-Laurent expansion of the corrections to the Schwarzschild potentials}), at this stage we get
\begin{equation}
    \begin{aligned}
        &\nu(r) = \ln\left(1-\frac{2MG}{r}\right) + \sum_{i=2}^{n-1} \frac{\alpha_i}{r^i} - (n+1)\frac{c_2}{r^n} + \mathcal{O}\left(\frac{1}{r^{n+1}}\right),\\
        &\mu(r) = -\ln\left(1-\frac{2MG}{r}\right) - \sum_{i=2}^{n-1} \frac{\alpha_i}{r^i} - (n+1)\frac{c_3}{r^n}+ \mathcal{O}\left(\frac{1}{r^{n+1}}\right),\\
        &c_2+c_3 = c_1,\quad n\geq 2,
    \end{aligned}
    \label{eq: general expression of mu and nu up to 1/r^n after integrating the first EE}
\end{equation}
where the conditions (\ref{eq: conditions on the functions of the metric in the quasi Schwarzschild solution}) have also been imposed. Notice that given the structures (\ref{eq: leading order correction to the derivative of f in the coordinates}) and (\ref{eq: general expression of mu and nu up to 1/r^n after integrating the first EE}), the modified EE and the expressions of the curvature tensors can be trusted up to order $\mathcal{O}\left(\frac{1}{r^{n+2}}\right)$, since both are second order in the derivatives of the metric.

For brevity, let us write $\nu$ as 
\begin{equation}
    \nu = s + \alpha - (n+1)\frac{c_2}{r^n}, 
\end{equation}
where $s$ is the usual Schwarzschild contribution and $\alpha$ is the sum of the terms $\frac{\alpha_i}{r^i}$, $i = 2,\dots,n-1$. Equation (\ref{eq: proper complete set of modified Einstein equations in f(R) with spherical symmetry - 2}) gives 
\begin{multline}
    e^s(\alpha_{rr} + \alpha_r^2 + 2s_r \alpha_r) + \frac{2}{r^2}\left(e^{-\alpha} - 1\right) \\
    +\frac{1}{2}(n^2+n-2)\frac{1}{r^{n+2}}[c_1 - (n+1)c_2] = 0,
    \label{eq: second modified EE evaluated with mu and nu with the subleading terms wrt Sch up to 1/r^n}
\end{multline}
where Eq.\ (\ref{eq: formula for mu + nu to order 1/r^n}) has been used together with the condition
\begin{equation*}
    \frac{1}{r^2}\left(e^{-s} -1\right) + \frac{s_{rr}}{2} + \frac{s_r^2}{2} = 0
\end{equation*}
holding for the Schwarzschild potential. Equation (\ref{eq: second modified EE evaluated with mu and nu with the subleading terms wrt Sch up to 1/r^n}) should be satisfied order by order in the inverse powers of the radius. Inserting the expansion of $\alpha$ one can realize that, starting from $k = 2$, every order $\mathcal{O}\left(\frac{1}{r^{k+2}}\right)$, $k = 2,\dots,n-1$ gives the condition 
\begin{equation}
    \frac{\alpha_k}{r^{k+2}}[k(k+1)-2] = 0,
\end{equation}
which is satisfied only if $\alpha_k = 0$. This implies the vanishing of every $\frac{\alpha_i}{r^i}$ term in  (\ref{eq: general expression of mu and nu up to 1/r^n after integrating the first EE}). 

The second line in (\ref{eq: second modified EE evaluated with mu and nu with the subleading terms wrt Sch up to 1/r^n}) instead implies $c_2 = \frac{1}{n+1}c_1$ and, together with (\ref{eq: general expression of mu and nu up to 1/r^n after integrating the first EE}), $c_3 = \frac{n}{n+1}c_1$. The leading correction to the Schwarzschild potentials is finally
\begin{equation}
    \begin{aligned}
        &\phi(r) = 1 + \frac{c_1}{r^n} +\mathcal{O}\left(\frac{1}{r^{n+1}}\right) ,\\
        &\nu(r) =  \ln\left(1-\frac{2MG}{r}\right) - \frac{c_1}{r^n} +\mathcal{O}\left(\frac{1}{r^{n+1}}\right),\\
        &\mu(r) = -  \ln\left(1-\frac{2MG}{r}\right) - n\frac{c_1}{r^n} +\mathcal{O}\left(\frac{1}{r^{n+1}}\right),\\
        &n\geq 2.
        \label{eq:Leading order corrections to the scalar and the Schwarzschild potentials}
    \end{aligned}
\end{equation}
Notice that $\mu$ and $\nu$ can also be written in a more familiar form 
\begin{equation}
\begin{aligned}
    &\nu(r) = \ln\left(1-\frac{2MG}{r} -\frac{c_1}{r^n}\right),\\
    &\mu(r) =  - \ln\left(1-\frac{2MG}{r} +n\frac{c_1}{r^n}\right),
\end{aligned}
\end{equation}
where it is left intended that the expressions hold up to $\mathcal{O}\left(\frac{1}{r^{n+1}}\right)$. To avoid pedantry, from now on we explicitly indicate the presence of higher orders in the expressions only when the order may be nontrivial. 

Inserting in (\ref{eq: line element of a spherical system in proper coordinates}) the line element is obtained
\begin{multline}
ds^2 = - \left(1- \frac{2MG}{r} - \frac{c_1}{r^n}\right) dt^2 \\+ \frac{1}{1-\frac{2MG}{r} + n\frac{c_1}{r^n}} dr^2  +r^2d\Omega^2,
\label{eq: modified Schwarzschild metric in f(R) with spherical symmetry}
\end{multline}
or equivalently,
\begin{multline*}
    ds^2 = 
    - \left(1- \frac{2MG}{r}\right) dt^2  
    +\frac{1}{1-\frac{2MG}{r}} dr^2 + r^2d\Omega^2 \\
    - \frac{c_1}{r^n}(-dt^2 + ndr^2).
\end{multline*}
This is the usual Schwarzschild line element plus a linear correction in $(1-\phi)$. Notice that nothing can be said about a possible shift in the horizon, since this solution holds at radii much bigger than the Schwarzschild radius. 

Last, Eq.\ (\ref{eq: proper complete set of modified Einstein equations in f(R) with spherical symmetry - 4}) implies
\begin{equation}
    f(R(r)) = 3n(n-1)\frac{c_1}{r^{n+2}} + \mathcal{O}\left(\frac{1}{r^{n+3}}\right).
    \label{eq: f(R) in coordinates calculated from the modified EE}
\end{equation}

\subsection{Recovering \texorpdfstring{$f(R)$}{f(R)}}

Equation (\ref{eq: f(R) in coordinates calculated from the modified EE}) can be understood computing the leading correction to the scalar curvature of the Schwarzschild solution. Using (\ref{eq: Ricci scalar for the metric of a spherical system}) we find 
\begin{equation}
    R = 3n (n-1)\frac{c_1}{r^{n+2}} + \mathcal{O}\left(\frac{1
    }{r^{n+3}}\right),
    \label{eq: leading correction to the scalar curvature computed from the metric}
\end{equation}
and hence, equation (\ref{eq: f(R) in coordinates calculated from the modified EE}) just tells us that 
\begin{equation}
    f(R) = R + \text{higher orders}, 
\end{equation}
again confirming the compatibility with GR in the weak field region.

Because of the structure of the curvature tensors, evidently the leading correction to the Einstein-Hilbert action cannot be recovered from the modified EE. However, it can be done using the definition of $\phi(R(r)) \equiv\pd{f(R)}{R} = 1 + \frac{c_1}{r^n}$ and Eq.\ (\ref{eq: leading correction to the scalar curvature computed from the metric}) in 
\begin{multline}
    f(R(r)) = \int \phi(R(r)) \frac{dR}{dr}dr \\ 
    = 3n(n-1)\frac{c_1}{r^{n+2}} + 3n(n-1)\frac{n+2}{2n+2}\frac{c_1^2}{r^{2n+2}} + \text{const}. 
    \label{eq: f(R) as integration in the coordinates of phi(R(r))}
\end{multline}
Inverting $R(r)$ in (\ref{eq: leading correction to the scalar curvature computed from the metric}) we finally get the $f$ as a function of the scalar curvature
\begin{equation}
    f(R) = R + \frac{1}{2}|c_1|^{\frac{2}{n+2}}\frac{n+2}{(n+1)(3n^2-3n)^{\frac{n}{n+2}}}|R|^{2\frac{n+1}{n+2}}.
    \label{eq: a posteriori recovered form of the leading correction to Einstein-Hilbert action}
\end{equation}
The constant of integration is set to $0$ asking that GR be recovered when $c_1 = 0$. Some comments are in order. 

\subsubsection{How general is this \texorpdfstring{$f(R)$}{f(R)}?}
One may argue that (\ref{eq: a posteriori recovered form of the leading correction to Einstein-Hilbert action}) is actually the value of $f(R)$ on-shell for the solution (\ref{eq: modified Schwarzschild metric in f(R) with spherical symmetry}). In other words, naming the value in (\ref{eq: leading correction to the scalar curvature computed from the metric}) $R_{\text loc}$ one could argue that $\phi$ in (\ref{eq: f(R) as integration in the coordinates of phi(R(r))}) is really 
\begin{equation*}
    \phi_{loc} = \lim_{R\to R_{loc}} \phi(R)
\end{equation*}
and that the general $f(R)$ is obtained integrating the unknown $\phi(R)$.

This is only partially true, the reason being that the metric (\ref{eq: modified Schwarzschild metric in f(R) with spherical symmetry}) is not an exact solution; hence, (\ref{eq: f(R) as integration in the coordinates of phi(R(r))}) is not really exactly evaluated on-shell. To see this consider (\ref{eq: Schwarzschild potentials with arbitrary corrections}) for arbitrary functions $\varphi,g,m$. The modified EE outside a static, spherical source when linearized around those functions imply $g_{rr} = g_{rr}(\varphi_{rr}), m_{rr} = m_{rr}(\varphi_{rr})$ and therefore $R(r) \propto \phi_{rr}$. The function $f(R)$ can then be recovered as $f(R) = \int \phi \, d\phi_{rr}$. This is only because of spherical symmetry and the assumed strong compatibility with GR. The analytical form in (\ref{eq: leading correction to the scalar curvature computed from the metric}) is instead a consequence of the regularity we assumed for $\phi$ far from the source (\ref{eq: Taylor-Laurent expansion of the corrections to the Schwarzschild potentials}), (\ref{eq: Schwarzschild potentials with arbitrary corrections}), which nonetheless allows certain generality of the discussion. 

Keeping this in mind, our assessment of the generality of $f(R)$ in (\ref{eq: a posteriori recovered form of the leading correction to Einstein-Hilbert action}) is the following: it is the leading correction to the Einstein-Hilbert action in the low curvature limit, strictly speaking holding outside of a spherical, static source and assuming some regularity of the metric at infinity. One may try to generalize the use of this $f(R)$ as a \enquote{boundary} condition for the low curvature regimes in other contexts, but this requires an additional assumption, even though motivated by the study of spherical sources. 

\subsubsection{What are the next to leading corrections?}
We are now in the position to understand what the next to leading correction looks like. Considering 
\begin{equation}
    \phi = 1 + \frac{c_1}{r^n} + \frac{c_2}{r^{n+1}} + \mathcal{O}\left(\frac{1}{r^{n+2}}\right),
\end{equation}
where again $c_2 = [c_2]\,l^{n+1}$, quite generally leads to 
\begin{equation}
    R = \frac{\rho_1(c_1)}{r^{n+2}} + \frac{\rho_2(c_2)}{r^{n+3}} + \mathcal{O}\left(\frac{1}{r^{n+4}}\right).
    \label{eq: general form of the next to leading correction to the scalar curvature of the Scwarzschild solution}
\end{equation}
The term $\rho_1(c_1)$ is known from the leading order $\rho_1(c_1) = 3n(n-1)c_1$, while the term $\rho_2(c_2)$ can be computed along the same lines first solving the EE and then computing the scalar curvature. The $f(R)$ function is recovered from 
\begin{multline}
    f(R(r)) = \int \left(1 + \frac{c_1}{r^n} + \frac{c_2}{r^{n+1}}\right)\\ \times\left[-(n+2)\frac{\rho_1}{r^{n+3}} - (n+3)\frac{\rho_2}{r^{n+4}}\right]dr,
\end{multline}
where the integrand can be trusted up to $\mathcal{O}\left(\frac{1}{r^{2n+5}}\right)$. Integrating we get 
\begin{multline}
    f(R(r)) = \frac{\rho_1(c_1)}{r^{n+2}} + \frac{\rho_2(c_2)}{r^{n+3}} + \frac{1}{2} \frac{n+2}{n+1}\frac{\rho_1c_1}{r^{2n+2}} \\+ \frac{1}{2n+3}[(n+2)\rho_1c_2 + (n+3)\rho_2c_1]\frac{1}{r^{2n+3}}.
    \label{eq: next to leading order correction to the Einstein-Hilbert action in the coordinates}
\end{multline}
Notice that further corrections to the scalar seem to spoil the perturbative expansion of the $f$, since they dominate the corrections from the higher powers of $R$. However, the first terms in the series always conspire to form the Einstein-Hilbert part of the action; hence, it is still well defined as an expansion in the scalar curvature. 

Having (\ref{eq: next to leading order correction to the Einstein-Hilbert action in the coordinates}) and the leading order correction (\ref{eq: a posteriori recovered form of the leading correction to Einstein-Hilbert action}), one can realize that the next to leading correction in the scalar curvature takes the form
\begin{equation}
    f(R) = R + \frac{1}{2}\frac{n+2}{n+1} c_1\rho_1 \left|\frac{R}{\rho_1}\right|^{2\frac{n+1}{n+2}} + \alpha(n,c_1,c_2) |R|^{\frac{2n+3}{n+2}}.
\end{equation}
The prefactor $\alpha$ can be fixed inserting the expression for $R$ (\ref{eq: general form of the next to leading correction to the scalar curvature of the Scwarzschild solution}), expanding to $\mathcal{O}\left(\frac{1}{r^{2n+4}}\right)$
\begin{equation}
    f(R) = R + \frac{1}{2}\frac{n+2}{n+1}\frac{c_1\rho_1}{r^{2n+2}} + \frac{1}{r^{2n+3}}\left(c_1\rho_2 + \alpha |\rho_1|^{\frac{2n+3}{n+2}}\right)
\end{equation}
and comparing with (\ref{eq: next to leading order correction to the Einstein-Hilbert action in the coordinates})
\begin{equation}
    \alpha = \frac{1}{|\rho_1|^{\frac{2n+3}{n+1}}}\left(\frac{n+2}{2n+3}\rho_1c_2 - \frac{n}{2n+3}\rho_2c_1\right).
\end{equation}

\subsubsection{What is the meaning of the coupling?}
As said, the dimensionate parameter $c_1$ should be intended as split in $c_1 = [c_1]\, l^n$. The fundamental length $l$ is needed to provide the correct dimensions in the gravitational action, and it determines the typical scale of deviation from GR. At this stage $l$ is a free parameter of the theory and may or may not be determined by the Newton constant. The numerical factor $[c_1]$ is instead related to the expansion of the exact $f$ in the low curvature limit. 

In the second part of the work we show how it is possible to set upper bounds on $c_1$ computing the corrections to the outcomes of the local tests and comparing with the observations. 
Clearly, as long as the leading order alone is considered in the expansion of $f$, one can never really distinguish $l$ and $[c_1]$. To this purpose one should either make a preliminary assumption on $[c_1]$ or consider the order next to leading.  

\subsubsection{Nonanalyticity} 
What is noteworthy about our procedure is that the resulting expression for the $f(R)$, Eq.\ (\ref{eq: a posteriori recovered form of the leading correction to Einstein-Hilbert action}), is nonanalytic at the background value $R = 0$. Several consequences could be traced back to this feature \cite{Capozziello2010}, as we are going to see. Here we stress that derivatives of second order or higher are not well defined in the limit $R\to 0$. 

More than that, consider what we have called the GR limit $c_1 \to 0$ for a theory defined by the action (\ref{eq: gravitational action of f(R)}), (\ref{eq: a posteriori recovered form of the leading correction to Einstein-Hilbert action}) in the low curvature regime. This is well defined for any derivative order only as long as $R$ stays different from zero, which would imply discarding vacuum GR solutions.\footnote{Notice that this problem does not occur if $f(R)$ is analytic.} Of course, this is not physically acceptable, and it is certainly not consistent with our perturbative procedure. 

Another formulation of the same problem is that the limit $c_1\to 0$ in the gravitational action does not commute with (higher than second order) derivation with respect to $R$ on $R= 0$ GR solutions. This means that, in general, we cannot smoothly recover GR results in observables (if any) involving second- or higher-order derivatives of the gravitational action. 

If we consider the family of local solutions with $R$ in (\ref{eq: leading correction to the scalar curvature computed from the metric}), as we should, the situation is even bleaker because the limit $c_1\to 0$ also automatically implies $R \to 0$. In fact, one can verify that the second derivative of our $f(R)$ is independent from $c_1$, while higher derivatives depend upon inverse powers of it. 

\subsubsection{Effective range of the scalar}
As said, a nonlinear function $f(R)$ in the gravitational action (\ref{eq: gravitational action of f(R)}) provides an additional scalar degree of freedom with respect to just the metric tensor. This is already clear from the field equations (\ref{eq: Einstein equations for f(R)}), since they involve (at most) second-order derivatives of the metric and of $\phi \equiv \pd{f}{R}$. 
Another way to see this\footnote{The most rigorous way to understand the degrees of freedom of $f(R)$ gravity, as in any covariant theory, would be by looking at its Hamiltonian formulation and inspecting the corresponding phase space \cite{Deruelle2009}} is by looking at the classical equivalence between $f(R)$ and a scalar tensor-theory \cite{Ohta2018}, which also sheds some light on the properties of the scalar.

Consider the action
\begin{equation}
    S = \frac{1}{2k}\int d^4x \sqrt{-g} \left[\frac{df(\chi)}{d\chi}(R-\chi) + f(\chi)\right],
    \label{eq: gravitational action with auxiliary field chi which is equivalent to the f(R) action}
\end{equation}
where $f$ is an arbitrary function, as in (\ref{eq: gravitational action of f(R)}), and $\chi$ is an additional scalar field. Variation with respect to $\chi$ and $g$ gives the field equations

\begin{align}
    &\delta_\chi S:\,\,\,\, \frac{d^2f(\chi)}{d\chi^2}(R-\chi) = 0,\\
    &\delta_g S:\,\,\,\, \frac{df(\chi)}{d\chi} R_{\mu\nu} - \frac{1}{2} g_{\mu\nu}\left[\frac{df(\chi)}{d\chi}(R-\chi) + f(\chi)\right] 
    \label{eq: g variation of gravitational action with auxiliary chi field which is equivalent to f(R)}
    \\&\nonumber\hspace{2,8cm}+ g_{\mu\nu} \nabla^2 \frac{df(\chi)}{d\chi} - \nabla_\mu\nabla_\nu \frac{df(\chi)}{d\chi}  = 0.
\end{align}

If $\frac{d^2f(\chi)}{d\chi^2}$ is regular and different from zero, that is, $f$ is nonlinear, then $\chi = R$ and Eq.\ (\ref{eq: g variation of gravitational action with auxiliary chi field which is equivalent to f(R)}) reduces to Eq.\ (\ref{eq: Einstein equations for f(R)}). Since classically the fields are always on-shell, we can say that if $f$ is nonlinear the field theory described by (\ref{eq: gravitational action with auxiliary field chi which is equivalent to the f(R) action}) is equivalent to the $f(R)$ theory deriving from (\ref{eq: gravitational action of f(R)}). Notice that the equations are now linear in the curvature and that $\chi$ is indeed a dynamical field. 

To render the action in a scalar-tensor form we need to redefine the scalar field introducing $\phi \equiv \frac{df(\chi)}{d\chi}$. If the equation is invertible, we can solve $\chi$ in terms of $\phi$, $\chi = \chi(\phi)$, so that the action becomes
\begin{equation}
    S = \frac{1}{2k}\int d^4x \sqrt{-g}[\phi(R-\chi(\phi)) + f(\chi(\phi))].
    \label{eq: gravitational action equivalent to the f(R) action with phi and g}
\end{equation}
Variation with respect to $\phi$ and $g$ yields
\begin{align}
    &\delta_\phi S:\,\,\,\, R - \chi(\phi) - \phi \frac{d\chi(\phi)}{d\phi} + \frac{df(\chi(\phi))}{d\phi} = 0,
    \label{eq: variation of phi-g gravitational action equiv to f(R) action with respect to phi}
    \\
    &\delta_g S:\,\,\,\, \phi R_{\mu\nu} - \frac{1}{2} g_{\mu\nu}\left[\phi(R-\chi(\phi)) + f(\chi(\phi))\right] 
    \label{eq: variation of phi-g the gravitational action equiv to f(R) action with respect to g}
    \\&\nonumber\hspace{2,8cm}+ g_{\mu\nu} \nabla^2 \phi - \nabla_\mu\nabla_\nu \phi  = 0.
\end{align}
Equation (\ref{eq: variation of phi-g gravitational action equiv to f(R) action with respect to phi}) gives $R = \chi(\phi)$, once the definition of $\phi$ is used, and inserting in (\ref{eq: variation of phi-g the gravitational action equiv to f(R) action with respect to g}) again Eq.\ (\ref{eq: gravitational action of f(R)}) is recovered. The equivalence with a scalar-tensor theory becomes clearer by introducing the potential
\begin{equation}
    V(\phi) = \phi \chi(\phi) - f(\chi(\phi)),\,\,\, \chi(\phi): \frac{df(\chi)}{d\chi} = \phi,
    \label{eq: potential for phi in the scalar tensor theory equivalent to f(R)}
\end{equation}
and rewriting the action (\ref{eq: gravitational action equivalent to the f(R) action with phi and g}) as 
\begin{equation}
    S = \frac{1}{2k}\int d^4x \sqrt{-g}[\phi R - V(\phi)].
\end{equation}
Taking the trace of Eq.\ (\ref{eq: variation of phi-g the gravitational action equiv to f(R) action with respect to g}) we see that the scalar $\phi$ obeys
\begin{equation}
    \nabla^2\phi - \frac{1}{3}\left(\phi \frac{dV(\phi)}{d\phi} - 2V(\phi)\right) = 0,
    \label{eq: Klein-Gordon equation satisfied by the scalar dof in f(R)}
\end{equation}
where $R = \frac{dV(\phi)}{d\phi}$ has been used. This is a Klein-Gordon equation for $\phi$, subject to the effective potential 
\begin{equation}
    V_{\text{eff}}(\phi):\,\,\, \frac{dV_{\text{eff}}(\phi)}{d\phi} =  \frac{1}{3}\left(\phi \frac{dV(\phi)}{d\phi} - 2V(\phi)\right).
\end{equation}

Now, if $V_{\text{eff}}(\phi)$ is bounded from below and admits one global minimum, there is a natural definition of the effective mass of $\phi$ for vacuum solutions near the minimum
\begin{equation}
    m^2_{eff} \equiv \left.\frac{d^2 V_{\text{eff}}(\phi)}{d\phi^2}\right|_{\phi_0},
    \label{eq: most reliable definition of the effective mass of the scalar dof introduced by f(R)}
\end{equation}
where $\phi_0$ is the minimum of $V_{\text{eff}}(\phi)$. This is a measure of the effective range of the $f(R)$ extension \cite{Capozziello2010,Faraoni2009}. We should mention that (at least) two other nonequivalent definitions are possible \cite{Faraoni2009}, one making use of $V(\phi)$ in (\ref{eq: potential for phi in the scalar tensor theory equivalent to f(R)}) instead of $V_{\text{eff}}(\phi)$ and one employing the Einstein frame representation of $V(\phi)$ \cite{Faraoni2009}. However, as argued in \cite{Faraoni2009} the most reliable definition is still the one given in (\ref{eq: most reliable definition of the effective mass of the scalar dof introduced by f(R)}), since it is directly related to a Klein-Gordon equation for $\phi$ (\ref{eq: Klein-Gordon equation satisfied by the scalar dof in f(R)}).

Let us consider the class of $f(R)$ extension 
\begin{equation}
    f(R) = R + \alpha |R|^k,\,\,\,\,\, \alpha > 0,\,\, k >1.
    \label{eq: f(R) class studied for the effective range}
\end{equation}
The function we have found in (\ref{eq: a posteriori recovered form of the leading correction to Einstein-Hilbert action}) belongs to this family with $\frac{3}{2}\leq k<2$, in the low curvature regime. 
The effective potential for such a class of functions is a straightforward computation 
\begin{equation}
    V_{eff}(\phi) = \frac{\alpha}{3}\frac{k-1}{2k-1}\left(\frac{|\phi-1|}{\alpha k}\right)^{\frac{k}{k-1}}[(2-k)\phi + 3(k-1)].
\end{equation}
For $k \neq 2$ the potential has a local minimum at $\phi = 1$ and a local maximum for $\phi = 2 \frac{k-1}{k-2}$, but its unbounded from below in the limit $\phi \to -\infty$, for $1<k<2$, and in the opposite limit for $k>2$. Therefore applying the definition of mass in these cases is only of limited value since the solutions are unstable. Moreover, the definition of a mass essentially relies on the possibility of Taylor expanding the potential around its minimum, but unless $\frac{k}{k-1}$ is an even integer, in which case the effective potential is analytical at $\phi = 1$, higher-order derivatives may not be defined or diverge and the expansion is not defined. Hence, it is not even clear if a consistent definition of a mass can be given this way. Interestingly, the case k= 2 corresponding to Starobinsky inflation is the only one in which the potential is bounded, with a global minimum at $\phi = 1$. The effective mass is thus well defined in this case and given by $m^2_{\text{eff}}|_{k=2} = \frac{1}{6\alpha}$.

Let us come to the definition of the effective range for our $f(R)$ in (\ref{eq: a posteriori recovered form of the leading correction to Einstein-Hilbert action}). As said, it belongs to the class (\ref{eq: f(R) class studied for the effective range}) with $\frac{3}{2}\leq k<2$ but only in the low curvature limit; hence, we cannot really trust the potential far away from $\phi = 1$ and certainly not its asymptotic divergences. In other words, it is possible that higher-order terms in the expansion of the $f(R)$ can stabilize the solutions. Other than that, at the moment we do not even have a clear figure of whether a global minimum can appear and be shifted from the GR local minimum when considering the exact $f(R)$. We must therefore conclude that such a definition of the effective range of the scalar cannot provide information in our case.

This being said, we will continue referring to a sort of range of the scalar interaction in a nonrigorous but intuitive way. In this context, the range is determined by the magnitude of $c_1$ in relation with the solar system typical scales.

\section{PPN parameters}
\label{sec:PPN parameters}
Before looking at the local tests we briefly discuss the prediction of the theory for the PPN parameters \cite{Misner2017}. 

\subsection{Isotropic metric}
It is conventional to study the metric (\ref{eq: most general line element for spherically symmetric systems}) for a spherical, static source in the isotropic form 
\begin{equation}
    ds^2 = -H(\rho)dt^2 + J(\rho)[d\rho^2 + \rho^2(d\theta^2 + \sin(\theta)^2 d\varphi^2)],
    \label{eq: spherically symmetric and static metric in the isotropic form}
\end{equation}
where $H(\rho) \equiv B(r(\rho)),\, J(\rho) \equiv \frac{r^2(\rho)}{\rho^2}$ and 
\begin{equation*}
    \rho =\frac{1}{2}\rho_0 \exp(\int_{r_0}^{r}\sqrt{A(r')}\frac{dr'}{r'}).
    \label{eq: isotropic radius}
\end{equation*}
In particular, for the Schwarzschild solution
\begin{equation}
    ds^2_{\text{S}} = -\left(\frac{1-\frac{MG}{2\rho}}{1+\frac{MG}{2\rho}}\right)^2 dt^2 + \left(1+\frac{MG}{2\rho}\right)^4(d\rho^2 + \rho^2d\Omega^2).
\end{equation}
\subsection{Post-Newtonian approximation}
The post-Newtonian approximation (to leading order) requires one to limit $g_{tt}$ to second order in $\epsilon^2 = \frac{MG}{\rho}$ and $g_{ij}$ to first order\footnote{This is required in order to recover the correct Newtonian limit of the geodesic equation for a test particle \cite{Misner2017}.} \cite{Misner2017}. Hence to post Newtonian accuracy 
\begin{multline}
    ds^2_{\text{S}} = -\left(1 - \frac{2MG}{\rho} + 2 \frac{(MG)^2}{\rho^2}\right)dt^2  \\ 
    + \left(1 + \frac{2MG}{\rho}\right)(d\rho^2 + \rho^2 d\Omega^2).
    \label{eq: Schwarzschild solution to leading post newtonian accuracy}
\end{multline}

The structure of (\ref{eq: Schwarzschild solution to leading post newtonian accuracy}) for the leading corrections to the Newtonian limit of the metric is, in fact, general and does not really depend on the use of GR. By asking asymptotic flatness and that the source be at rest at the origin of the reference frame, in any metric theory of gravity (reliable in the Newtonian limit) one obtains 
\begin{multline}
    ds^2 = - \left(1 - \frac{2MG}{\rho} + 2 \beta \frac{(MG)^2}{\rho^2}\right)dt^2 \\ + \left(1 + 2 \gamma \frac{MG}{\rho}\right)(d\rho^2 + \rho^2d\Omega^2),
\end{multline}
again to leading post-Newtonian order. The parameters $\beta$ and $\gamma$ are two real constants, respectively quantifying the nonlinear deviation from the Newtonian potential and the curvature of the $3$D hypersurfaces. These are two of the ten parameters (PPN parameters) characterizing the post-Newtonian approximation of a general metric \cite{Misner2017}, the other eight being constrained by symmetry in this case and the fact that the source is assumed at rest in the PPN frame. 

GR foresees the values $\beta = \gamma = 1$; hence, precise measurements of those parameters strongly constrain possible modifications of Einstein's theory. Current bounds are $\gamma -1 = (2.1 \pm 2.3)\times 10^{-5}$ \cite{Will2014}, from the time delay measurements of the Cassini spacecraft \cite{Bertotti2003}, and $\beta - 1 = (-4.1 \pm 7.8)\times 10^{-5}$\footnote{This is found assuming \textit{a priori} the Cassini bound on $\gamma$.} \cite{Will2014} from the perihelion precession of Mercury.

\subsection{PPN parameters for \texorpdfstring{$f(R)$}{f(R)}}
In isotropic coordinates the metric (\ref{eq: modified Schwarzschild metric in f(R) with spherical symmetry}) reads 
\begin{equation}
    ds^2 = ds^2_{\text{S}} - \frac{c_1}{\rho^n}\left(-dt^2 + nd\rho^2\right).
\end{equation}
If $n >2$ the deviation from the Schwarzschild metric does not contribute to post Newtonian order and the prediction for $\gamma,\beta$ is the same as in GR. If $n = 2$, which is the minimum admissible value, to PPN accuracy we get 
\begin{multline}
    ds^2 = - \left[1 - \frac{2MG}{\rho} + 2\left(1-\frac{c_1^{(2)}}{2(MG)^2}\right)\frac{(MG)^2}{\rho^2}\right]dt^2 \\ 
    + \left(1 + 2 \frac{MG}{\rho}\right)(d\rho^2 + \rho^2d\Omega^2).
\end{multline}
The prediction for the PPN parameters are therefore $\gamma = 1$ (see e.g.\ \cite{Toniato2021}) and $\beta = 1 - \frac{c_1^{(2)}}{2(MG)^2}$. While the Cassini bound is naturally saturated, the bound on $\beta$ would imply 
$|c_1|^{(2)}\lesssim 1.2\times 10^{-6} \text{ mm}^2$.

\section{Solar system tests}
\label{sec:solar system tests}
In this section we discuss the predictions of the $f(R)$ extension of GR found in Sec.\ \ref{sec:Asymptotically Schwarzschild solutions} for the outcomes of four different solar system experiments: the gravitational redshift, the deflection of light, the precession of closed orbits and the Shapiro delay \cite{Weinberg1972}. For each of them we compute the leading correction to the value predicted by GR; moreover, we set an upper bound on $c_1$ at varying $n$ by comparing the predictions with measures of the sunlight gravitational redshift \cite{GonzalezHernndez2020}. 

For brevity we write general formulas for the observables holding in any spherically symmetric metric (\ref{eq2: eq: line element of a spherical system in proper coordinates -2})
\begin{equation*}
    ds^2 = -B(r)dt^2 + A(r)dr^2 + r^2 d\Omega^2.
\end{equation*}
In our case 
\begin{equation*}
    \begin{aligned}
       B(r) = 1- \frac{2MG}{r} - \frac{c_1}{r^n},\\
       A(r) = \frac{1}{1-\frac{2MG}{r} + n\frac{c_1}{r^n}}.
    \end{aligned}
\end{equation*}

\subsection{Gravitational redshift}
Consider two observers $O_1, O_2$ located fixed at $r_1$, $r_2 > r_1$ from a spherical, static source and equal angular coordinates. If $O_1$ sends a light signal of frequency $\nu_1$ to $O_2$, the frequency detected by the latter will be
\begin{equation}
    \nu_2 = \sqrt{\frac{B(r_1)}{B(r_2)}}\nu_1. 
\end{equation}
The shift in the frequency can be quantified by 
\begin{equation}
    z \equiv \frac{\nu_1 - \nu_2}{\nu_1} = 1 -\sqrt{\frac{B(r_1)}{B(r_2)}} .
\end{equation}
In the case of metric (\ref{eq: modified Schwarzschild metric in f(R) with spherical symmetry}) we get a redshift 
\begin{equation}
    \nu_2 = \sqrt{\frac{1-\frac{2MG}{r_1}}{1-\frac{2MG}{r_2}}}\nu_1 - \frac{1}{2}\frac{c_1}{r_1^n}\left(1 - \frac{r_1^n}{r_2^n}\right)\nu_1,
\end{equation}
\begin{equation}
    z = 1 -\sqrt{\frac{1-\frac{2MG}{r_1}}{1-\frac{2MG}{r_2}}} + \frac{1}{2}\frac{c_1}{r_1^n}\left(1 - \frac{r_1^n}{r_2^n}\right).
\end{equation}
The first term of the equations above is the usual GR prediction, the second is the $f(R)$ contribution. Notice that the $f(R)$ extension enhances the magnitude of the redshift if $c_1 > 0$ and reduces it if $c_1 < 0$.

To set a bound on $c_1$ we use observations of the sunlight gravitational redshift performed by González Hernández \textit{et al}.\ in 2020 \cite{GonzalezHernndez2020}. By analyzing the shift in the spectral lines of Fe from the sunlight reflected by the moon, the authors found a mean global line shift of $z = \frac{1}{c}(638 \pm 6)$ m\,s$^{-1}$, which is consistent with the GR prediction\footnote{This value also accounts for the gravitational field of the Earth \cite{GonzalezHernndez2020}.} $\sim 633$ m\,s$^{-1}$. We can therefore take 
\begin{equation*}
    \frac{1}{2}\frac{|c_1|}{r_1^n}\left(1 - \frac{r_1^n}{r_2^n}\right) \lesssim 1 \frac{\text{m\,s}^{-1} }{c}.
\end{equation*}

The resulting upper bounds on $c_1$ (in units of the Schwarzschild radius of the Sun) for $n$ up to four are shown in Table \ref{tab:upper bounds on c_1 at varyin n from sunlight redshift measurements}.

\begin{table}[!ht]
    \centering
    \begin{tabular}{p{1cm} p{1.7cm}}
    \hline\hline
    $n$ &  $\sqrt[n]{\frac{|c_1|}{(2M_\odot G)^n}}$\\
    \hline
    $2$ & $19.14$ \\
    \hline
    $3$ & $441.24$ \T\\
    \hline
    $4$ & $2118.45$\T\\
    \hline\hline
    \end{tabular}
    \caption{Upper bounds on $c_1$ at varying $n$ from sunlight gravitational redshift measurements \cite{GonzalezHernndez2020}.}
    \label{tab:upper bounds on c_1 at varyin n from sunlight redshift measurements}
\end{table}

The length scale of the deviation from GR introduced by $f(R)$ increases with the order of the correction. Moreover, notice that the bound on $|c_1|^{(2)}$ found in the previous section from the precession of the perihelion of Mercury is way stronger, due to the greater accuracy of that measurement.  

\subsection{Gravitational lensing}
Consider light reaching the Earth from a very distant star.  Because of the presence of the Sun, the geodesic of a photon will not be a straight line in the Euclidean sense. 

Let us orient the reference frame so that the motion of the photon lies on the plane $\theta = \frac{\pi}{2}$ and the incident direction is at $\varphi  = 0$.
Inspecting the null geodesic equation \cite{Weinberg1972} the trajectory is found as
\begin{equation}
    \varphi(r)= \int_r^\infty 
    \sqrt{A(r')}\left[\lrb\frac{r'}{r_0}\rrb^2\frac{B(r_0)}{B(r')}-1\right]^{-\frac{1}{2}}\frac{dr'}{r'}.
    \label{eq: deflection angle formula for a spherically symmetric metric}
\end{equation}
Here $r_0$ is the distance of closest approach to the Sun. Notice that this solution refers to the branch in which $r$ decreases from infinity to $r_0$, the other branch being $\varphi'(r) = 2\varphi(r_0) - \varphi(r)$. For the solution (\ref{eq: modified Schwarzschild metric in f(R) with spherical symmetry}) and adopting the same perturbative spirit we can expand the integrand up to $\frac{1}{r}\times \frac{c_1}{r^n}$ and find the leading correction to the exact Schwarzschild result 
\begin{equation}
    \varphi(r) = \varphi(r)|_{\text{Schwarzschild}} + \varphi(r)|_{f(R)},
    \label{eq: formal decomposition of the azimuthal angle in Schwarzschild + f(R) contributions}
\end{equation}
\begin{multline}
   \varphi(r)|_{f(R)} = -\frac{1}{2}c_1\int_r^{+\infty}\frac{dr}{r} \\\times \frac{1}{\sqrt{\left(\frac{r}{r_0}\right)^2-1}}\left[\frac{n}{r^n} + \frac{r^2}{r^2-r_0^2}\left(\frac{1}{r^n} - \frac{1}{r_0^n}\right)\right].
   \label{eq: integral for the correction to the deflection angle at some r in f(R)}
\end{multline}
Upon performing the change of variable $r \to y = \frac{r_0}{r}$ the integral can be performed analytically and gives 
\begin{equation}
    \varphi(r)|_{f(R)} =\, \frac{1}{2}\frac{c_1}{r_0^n}\left(\frac{r_0}{r}\right) \frac{1- \left(\frac{r_0}{r}\right)^n}{\sqrt{1 - \left(\frac{r_0}{r}\right)^2}}.
    \label{eq:f(R) contribution to the deflection angle of light by a spherical source}
\end{equation}
This is the leading contribution to the azimuthal trajectory of a photon coming from the $f(R)$ extension of GR. Usually also the Schwarzschild contribution to $\varphi(r)$ is computed perturbatively in $\frac{MG}{r}$ \cite{Weinberg1972, Misner2017}. Therefore, once an order $n$ is specified, in order to set an upper bound on $c_1$ one should first compute all the dominant GR contributions.

The final direction of a photon departing from $r_0$ again to infinity is $\lim\limits_{r\to\infty}\varphi'(r) = 2\varphi(r_0)$. In the Euclidean space this angle would amount to $\pi$; hence, the total deflection induced by the Sun is $\delta\alpha_\infty = 2\varphi(r_0) - \pi$. Inserting equations (\ref{eq: formal decomposition of the azimuthal angle in Schwarzschild + f(R) contributions}) and (\ref{eq:f(R) contribution to the deflection angle of light by a spherical source}) one can verify that there is no contribution to the total deflection angle from the $f(R)$ extension at order $\frac{c_1}{r_0^n}$, that is, 
\begin{equation}
    \delta\alpha_\infty = 2\varphi(r_0)|_{\text{Schwarzschild}} - \pi + \mathcal{O}(\frac{1}{r^{n+1}}).
\end{equation}
One should possibly consider the next to leading order in (\ref{eq:Leading order corrections to the scalar and the Schwarzschild potentials}) and perform again the computation. 

This being said, the quantity of interest for an astronomer on Earth is not the total deflection angle but rather the deflection suffered by the light at the point of its encounter with the telescope.\footnote{Even more interesting is the change in the relative angular separation between two stars, as their light passes by the Sun, or between the star of interest and another reference source whose light stays possibly unscattered.} Let us call $\alpha$ the angular separation between the Sun and the apparent direction of the star as seen from Earth. This is linked to the Earth position ($\varphi_{\text{E}}$) by 
\begin{equation}
    \alpha = \pi - \varphi_\text{E} + \delta\alpha, 
\end{equation}
where $\delta\alpha$ is the deflection we are interested in. Moreover, 
\begin{equation}
    \tan({\alpha}) = \frac{|u^\varphi|_E}{|u^r|_E} = \left.\frac{\sqrt{g_{\varphi\varphi}}d\varphi/d\lambda}{\sqrt{g_{rr}}dr/d\lambda}\right|_E = \left.\frac{r}{\sqrt{A(r)}}\frac{d\varphi}{dr}\right|_E,
\end{equation}
where $u^\varphi$ and $u^r$ are tangent vector fields to the trajectory of the photon pointing, respectively, in the azimuthal and the radial directions. The subscript indicates that the quantities must be evaluated at the Earth position. Inserting (\ref{eq: deflection angle formula for a spherically symmetric metric}) we get
\begin{equation}
    \tan(\pi - \varphi_E + \delta \alpha) = - \frac{1}{\sqrt{\lrb\frac{r_E}{r_0}\rrb^2\frac{B(r_0)}{B(r_E)}-1}}.
    \label{eq: relation between the tan of the apparent angle between sun and star and the metric functions}
\end{equation}
This equation can be solved iteratively by expanding the left-hand side (lhs) in $\delta\alpha$ and the right-hand side (rhs) in $\frac{MG}{r}$. Here we are interested in the correction introduced by the $f(R)$ extension up to order $\frac{c_1}{r^n}$. This can only appear (at most) at the $n$th order of $\delta \alpha$ when the lhs is expanded linearly in $\delta \alpha$.\footnote{Recall that solution (\ref{eq: modified Schwarzschild metric in f(R) with spherical symmetry}) holds up to terms $\mathcal{O}\left(\frac{1}{r^{n+1}}\right)$} Therefore, using 
\begin{equation*}
    \tan(\pi - \varphi_E + \delta \alpha) = - \tan(\varphi_E) + \frac{\delta\alpha}{\cos^2(\varphi_E)} + o(\delta\alpha^2)
\end{equation*}
and expanding the rhs of (\ref{eq: relation between the tan of the apparent angle between sun and star and the metric functions}) up to $\frac{c_1}{r^n}$ we deduce the correction
\begin{equation}
    \delta\alpha|_{f(R)} = \frac{1}{2}\frac{c_1}{r_0^n}\frac{\sin(\alpha)}{\cos(\alpha)}(1-\sin^n(\alpha)),
    \label{eq:correction to the deflection angle of the light from a distant star detected by an observer on Earth}
\end{equation}
where we have also used that $\alpha = \pi - \varphi_E + o(\frac{MG}{r})$. For a light ray grazing the Sun's limb ($\alpha = 0$) the correction vanishes, as found before. The correction also correctly vanishes when the ray comes opposite to the Sun's direction ($\alpha = \pi$). Surprisingly enough, it also vanishes when the light comes perpendicularly to the Sun's direction ($\alpha = \frac{\pi}{2}$).

Again, once an order $n$ is chosen and the dominant GR corrections are computed, one could use the result (\ref{eq:correction to the deflection angle of the light from a distant star detected by an observer on Earth}) to set a bound on $c_1$.

\subsection{Shapiro delay}
Here we consider light emitted from a source located at the Schwarzschild coordinates $r = r_1, \theta = \frac{\pi}{2}, \varphi = \varphi_1$. The coordinate time interval the photon takes to reach its minimum distance from the Sun ($r_0$) is \cite{Weinberg1972}
\begin{equation}
    \Delta t_{(1,0)} \equiv \Delta t(r_1,r_0) = \int_{r_1}^{r_0} dr \sqrt{\frac{A(r)}{B(r)}}\left[1 - \frac{B(r)}{B(r_0)}\left(\frac{r_0}{r}\right)^2\right]^{-\frac{1}{2}}
    \label{eq: general expression wit A and B for the time interval of a photon passing by a spherically symmetric static source}
\end{equation}
and suffers corrections in $\frac{M_\odot G}{r_0}$, with respect to the Euclidean value $\sqrt{r_1^2-r_0^2}$.

As before, we are interested in the correction produced by the $f(R)$ extension of GR. To this extent we expand the integrand in (\ref{eq: general expression wit A and B for the time interval of a photon passing by a spherically symmetric static source}) up to order $\mathcal{O}\left(\frac{1}{r^n}\right)$ and isolating the $f(R)$ contribution we find 
\begin{multline}
    \Delta t (r_1,r_0)|_{f(R)} = \frac{c_1}{r_0^n} \int_{r_1}^{{r_0}}dr \frac{1}{\sqrt{1-\left(\frac{r_0}{r}\right)^2}}\\\times\left\{- \frac{n-1}{2}\left(\frac{r_0}{r}\right)^n + \frac{1}{2}\left[1- \left(\frac{r_0}{r}\right)^n\right]\frac{r_0^2}{r^2-r_0^2}\right\}.
\end{multline}
Again performing the change $y = \frac{r_0}{r}$ the integral can be computed explicitly and gives 
\begin{equation}
    \Delta t (r_1,r_0)|_{f(R)} =  
    \frac{1}{2}\frac{c_1}{r_0^{n-1}} 
    \left[\frac{r_0}{r_1} \frac{1- \left(\frac{r_0}{r_1}\right)^{n-2}}{\sqrt{1-\left(\frac{r_0}{r_1}\right)^2}}\right].
\end{equation}
Notice that for $n=2$ there is no correction at this order.

The time of flight the photon takes to reach a target located at $r= r_2, \theta = \frac{\pi}{2}, \varphi = \varphi_2$ is $\Delta t_{1,2} = \Delta t_{1,0} \pm \Delta t_{2,0}$, with a plus sign if the minimum distance $r_0$ is crossed and a minus sign otherwise.

Let us consider an observer on Earth launching a signal toward a target in the solar system (a planet or a satellite, for example \cite{Will2014}) and measuring the time interval it takes to bounce back. The proper time elapsed for an observer on Earth when the photon reaches the target is $\Delta\tau_E(r_E,r_2) = \sqrt{g_{00}}|_E\Delta t(r_E,r_2) = \left(\sqrt{1 
- \frac{2M_\odot G}{r_E}} - \frac{1}{2}\frac{c_1}{r_E^n}\right)\Delta t(r_E,r_2)$. The total time elapsed when the signal again reaches Earth is twice as much. In particular, the $f(R)$ contribution to $\Delta\tau_E(r_E,r_0)$ is
\begin{equation}
    \Delta\tau_E(r_E,r_0)|_{f(R)} = \Delta t(r_E,r_0) - \frac{1}{2}\frac{c_1}{r_E^n}\sqrt{r_E^2 - r_0^2}.
\end{equation}
The $f(R)$ contribution to the total time of flight as measured by a clock on Earth finally is  
\begin{multline}
    2\Delta \tau(r_E, r_2)|_{f(R)} = \frac{c_1}{r_0^{n-1}}\frac{r_0}{r_E}\times\\ \times\left[\frac{1 - 2 \left(\frac{r_0}{r_E}\right)^{n-2} + \left(\frac{r_0}{r_E}\right)^n}{\sqrt{1-\left(\frac{r_0}{r_E}\right)^2}}\pm  (r_E \to r_2)
    \right].
\end{multline}

\subsection{Precession of closed orbits}
As a last application we consider the effect of the $f(R)$ extension on the precession of closed orbits. 

Consider a test massive particle in a bound orbit around the Sun. Let us call $a$ and $b$, respectively, the radial distances of the perihelion and the aphelion of its trajectory and orient the reference frame (at some coordinate time) so that $\varphi(a) = 0$. Starting from the perihelion, the angle swept by the particle when it reaches an intermediate radial position $a<r<b$ is given by
\cite{Weinberg1972}
\begin{multline}
    \varphi(r) = \int_a^r \frac{dr'}{r'}\frac{\sqrt{A(r')}}{r'} \\ \times \left[\frac{a^2(B^{-1}(r') - B^{-1}(a)) - b^2(B^{-1}(r') - B^{-1}(b))}{a^2b^2(B^{-1}(b) - B^{-1}(a))} - \frac{1}{r^2}\right]^{-\frac{1}{2}}.
    \label{eq: general expression for the angle swept by a massive particle in a bound orbit around the static, spherical source at some a<r<b}
\end{multline}
By symmetry, the total angle swept in a revolution around the Sun is $2\varphi(b)$; hence, $\Delta\varphi \equiv 2\varphi(b)- 2\pi$ quantifies the precession of the perihelion.

The procedure at this point goes along the same lines as for the other tests: we expand the integrand and isolate the $f(R)$ contribution. The result is 
\begin{multline}
    \varphi(b)|_{f(R)} = -\frac{1}{2}\frac{a}{2M_\odot G}\frac{c_1}{ a^{n}} \\ \times\int_a^b \frac{dr}{r}\left[\frac{r}{b} \frac{r-a}{b-a} + (a \longleftrightarrow b) - 1\right]^{-\frac{3}{2}}\\\times
    \left[\frac{r^2}{b(b-a)}\left(1 - \frac{a^n}{r^n}\right) - \frac{r(r-a)}{(b-a)^2}\left(\frac{1}{a^{n-1}}- \frac{a}{b^n}\right)\right. \\+\left. (a\longleftrightarrow b)\vphantom{\frac{r^2}{r^2}}\right].
    \label{eq:integral for f(R) correction to the precession of the perihelion}
\end{multline}
The integral is somewhat more complicated than in the other cases but can nonetheless be performed analytically, the important steps are reported in Appendix \ref{appendix: precession integral}. Here we give the result which can be written compactly as 
\begin{multline}
    \varphi(b)|_{f(R)} = \frac{\pi}{4M_\odot G}\frac{c_1}{a^{n-1}}(1+k)\\ \times \sum_{j = 0}^{n-2}\binom{n}{j+2}\frac{(2j+1)!}{(j!)^2}\left(\frac{k-1}{4}\right)^j.
    \label{eq: final formula for the leading correction introduced by f(R) to the angular precession of a closed orbit}
\end{multline}
where $k = \frac{a}{b}$. 
Notice that for $n = 2$ the result simplifies to 
\begin{equation}
    \varphi(b)|_{f(R), n =2} = \frac{\pi}{4M_\odot G} \frac{c^{(2)}_1}{a}\left(1+\frac{a}{b}\right). 
\end{equation}
The same conclusions holding for the other tests hold here about the possibility of setting bounds on $c_1$. Also, one could use the results already found in table \ref{tab:upper bounds on c_1 at varyin n from sunlight redshift measurements} from the gravitational redshift to infer concrete predictions on all the three tests considered in this section and compare with the experiments.

\section{Cosmology}
\label{sec:cosmology}
We began this work by motivating an $f(R)$ extension of GR as a possible answer to the shortcomings of the $\Lambda$-CDM model. Although a thorough discussion lies outside the scopes of the current work, to cosmology we shall now return. 

A detailed and very general account of the conditions for the cosmological viability of $f(R)$ models can be found in \cite{Amendola2007Conditions}. Based on those results, if we try to just apply the extension (\ref{eq: a posteriori recovered form of the leading correction to Einstein-Hilbert action}) in cosmology we fail to reproduce the late times inflation, in favor of a stable matter epoch. This failure is, of course, expected: the expression in (\ref{eq: a posteriori recovered form of the leading correction to Einstein-Hilbert action}) must be understood as the result of a weak field limit of an exact $f(R)$, strictly speaking outside a spherical, static source, and cannot be simply applied in cosmology. 

As an example of how things could get interesting, suppose that in (\ref{eq: general expression of mu and nu up to 1/r^n after integrating the first EE}) we ask for a more general Schwarzschild$-$de Sitter background as a boundary for our potentials, instead of just the Schwarzschild solution. The computation of the leading correction to the background metric is identical, and we get
\begin{multline}
    ds^2 = 
    - \left(1- \frac{2MG}{r} - \frac{\Lambda}{3}r^2\right) dt^2 + 
    \frac{1}{1-\frac{2MG}{r} - \frac{\Lambda}{3}r^2} dr^2 \\+ r^2d\Omega^2 
    - \frac{c_1}{r^n}(-dt^2 + ndr^2).
    \label{eq: correction to the Schwarzschild-de Sitter metric induced bby f(R)}
\end{multline}
Again, from the scalar curvature of the metric 
\begin{equation}
    R = 4\Lambda + 3n(n-1)\frac{c_1}{r^{n+2}}
\end{equation}
the corresponding $f(R)$ is recovered,
\begin{equation}
    f(R) = R - 2\Lambda + \frac{1}{2}|c_1|^{\frac{2}{n+2}}\frac{n+2}{(n+1)(3n^2-3n)^{\frac{n}{n+2}}}(|R-4\Lambda|)^{2\frac{n+1}{n+2}}.
    \label{eq: a posteriori recovered form of the leading correction to Einstein-Hilbert action with CC}
\end{equation}
The constant of integration is set so that the dominant term is the usual Einstein-Hilbert. As before, expression (\ref{eq: a posteriori recovered form of the leading correction to Einstein-Hilbert action with CC}) should be understood as the expansion of $f(R)$ in the limit $R\to 4\Lambda$, holding outside a spherical source and in the presence of a cosmological constant. 

This time the late times acceleration comes naturally, in fact we asked for it, but it is not clear if a standard matter epoch is reproduced and a numerical study would be required \cite{Amendola2007Conditions}. In passing, this example also shows the impact of the boundary conditions (which essentially determine the physical context we move in) we impose on the potentials (\ref{eq: Schwarzschild potentials with arbitrary corrections}).

Nonetheless, the direct application of the results found here is of fairly limited value because of their highly special intent (that is, the physics of the solar system). A more serious attempt in the cosmological context would be to start from scratch and apply the same ideology employed here: study the EE after the symmetry reduction and without any preliminary assumption on $f(R)$; impose the known dynamics of the Universe in different epochs as a strong boundary condition on the unknown functions of the metric and consider the most general but compatible class of corrections; make minimal regularity assumptions on the corrections, so that they can be found explicitly; and recover \textit{a posteriori} the $f(R)$. 
The work done here should serve as a consistency condition on the $f(R)$ and expression (\ref{eq: a posteriori recovered form of the leading correction to Einstein-Hilbert action}) should be recovered at the solar system scales (neglecting the acceleration). Eventually, one should perform concrete predictions for the relevant observables and constrain the free parameters by comparison with the observations. 

If this can be done and, more importantly, it shows to at least alleviate the problems of standard cosmology, then we could say that such a class of extensions of GR is a viable alternative. This will be the theme of a subsequent work.

\section{Conclusions}
The aim of this work was to build a class of $f(R)$ extensions of GR fully compatible with solar system tests. The key idea we wanted to implement (and to the best knowledge of the authors also the true novelty) was that the request of full compatibility with GR in the weak field limit could be enough to constrain the form of $f(R)$ and allow for explicit calculations. Therefore, without making any preliminary assumption on the mathematical properties of $f(R)$, we considered the modified EE around a spherical, static source (\ref{eq: proper complete set of modified Einstein equations in f(R) with spherical symmetry - 1})$-$(\ref{eq: proper complete set of modified Einstein equations in f(R) with spherical symmetry - 3}) and looked for solutions of the form (\ref{eq: Schwarzschild potentials with arbitrary corrections}) and (\ref{eq: conditions on the functions of the metric in the quasi Schwarzschild solution}). We banked on the fact that the relevant length scale of the system is the Schwarzschild radius of the source, which for a typical star like the Sun is much smaller than its visible radius. Hence, the surrounding gravitational field can always be considered in the weak field regime, where a strong compatibility with GR holds. We stress that considering a spherically symmetric, static setting and using a perturbative approach are not in and of themselves mandatory elements to achieve realistic $f(R)$'s. However, in the spirit of the principle of compatibility with GR at local scales and the idea of a bottom-up approach to $f(R)$, these appear as natural conceptual tools to employ in a first attempt in this direction.

To go further and perform explicit computations we needed to assume some regularity of the functions correcting the Schwarzschild potentials. In particular, we asked that these could be expanded in a Laurent series around the origin of the coordinates\footnote{This assumption discarded Yukawa-like corrections which are well known in the literature \cite{Capozziello2010}.} (as, in fact, the Schwarzschild potentials themselves are), allowing for a perturbative approach [Eq.\ (\ref{eq: leading order correction to the derivative of f in the coordinates} is emblematic on this point]. Although this is a simplifying choice which in principle restricts the class of possible $f(R)$'s, it still allows quite some generality in the discussion. The resulting solution to the modified EE at leading order in the corrections is displayed in (\ref{eq: modified Schwarzschild metric in f(R) with spherical symmetry}). 

Having the metric tensor made it possible to recover \textit{a posteriori} the corresponding $f(R)$ by integrating the scalar curvature, Eq.\ (\ref{eq: a posteriori recovered form of the leading correction to Einstein-Hilbert action}). This is the most delicate point of the discussion and deserves attention, because although it is true that the $f(R)$ computed this way is \enquote{on-shell} with respect to (\ref{eq: modified Schwarzschild metric in f(R) with spherical symmetry}), the latter encompasses a whole class of solutions in the weak field limit (which in our perspective are also the relevant ones). Therefore the expression (\ref{eq: a posteriori recovered form of the leading correction to Einstein-Hilbert action}) should really be understood as the small curvature limit of the $f(R)$ outside a spherical, static source. The expression of the leading correction to the Einstein-Hilbert action (\ref{eq: a posteriori recovered form of the leading correction to Einstein-Hilbert action}) and that of the leading correction to the Schwarzschild solution (\ref{eq: modified Schwarzschild metric in f(R) with spherical symmetry}) are the most important conceptual results of this paper. 

In the second part of the work we turned our attention to the free parameters in the solution (\ref{eq: modified Schwarzschild metric in f(R) with spherical symmetry}). As said, while $n$ determines the order of the corrections and lets us \enquote{move} within the $f(R)$ class, $c_1$ is related to some fundamental length scale, essentially determining the range of the scalar interaction introduced by the $f(R)$ extension \cite{Faraoni2009}. We first computed the PPN parameters $\gamma,\beta$ of the solution and found that, while the value of $\gamma$ agrees with GR for every $n$, the value of $\beta$ agrees with GR for $n >2$. For $n = 2$, which is the minimum admissible value of $n$, the experimental bounds on $\beta$ set the first upper bound $c_1^{(2)}\lesssim 1.2\times 10^{-6} \text{\,mm}^2$.

We went on studying the physical implications of such an extension of GR to the solar system tests, the idea being that the fundamental constant $c_1$ can be properly set by comparison with the experimental results \cite{Will2014}. We computed the leading corrections, with respect to the GR results, to the gravitational redshift, the gravitational lensing, the Shapiro delay, and the precession of closed orbits. In particular, comparison with the gravitational redshift of sunlight \cite{GonzalezHernndez2020} allowed us to place bounds on $c_1$ at varying $n$, the first three values are shown in Table \ref{tab:upper bounds on c_1 at varyin n from sunlight redshift measurements}. This showed that the typical length scale of the scalar interaction is not necessarily smaller than the Schwarzschild radius of a typical star and that it increases with the order $n$ of the correction. In principle, once the dominant GR contributions are computed, also the other three classical tests could be used to place bounds on $c_1$. On the other hand, the results already found in Table \ref{tab:upper bounds on c_1 at varyin n from sunlight redshift measurements} could be used to infer tentative predictions of the $f(R)$ contributions.

Eventually we came back to cosmology, the logical next step of this work. We argued that a direct application of the expression (\ref{eq: a posteriori recovered form of the leading correction to Einstein-Hilbert action}) is fundamentally wrong, although interesting suggestions can be found even so [see Eqs.\ (\ref{eq: correction to the Schwarzschild-de Sitter metric induced bby f(R)})$-$(\ref{eq: a posteriori recovered form of the leading correction to Einstein-Hilbert action with CC})]. We should instead apply the point of view adopted here and the entire methodology at the cosmological scales, asking compatibility with the known dynamics. The results presented here would serve as additional consistency conditions when local scales are reached. 

\begin{acknowledgments}
F.\ S.\ whishes to thank G.\ Bianchi, M.\ Fontana, and A.\ Massidda for stimulating discussions and useful comments on the manuscript. 
F.\ S.\ also thanks M.\ Rossati for her collaboration during the preliminary part of the work. O.\ F.\ P.\ and F.\ S.\ are grateful to S.\ L.\ Cacciatori for his suggestions and for his careful revision which has substantially improved the manuscript.
\end{acknowledgments}

\appendix
\section{SPHERICALLY SYMMETRIC METRIC}
\label{appendix: Spherically symmetric metric}
We are interested in the gravitational field generated by a spherical source such as a star. The most general metric for such systems can be written in spherical coordinates as 
\begin{equation}
    ds^2 = -B(r,t)dt^2 + A(r,t)dr^2 + C(r,t)drdt + D(r,t)d\Omega^2,
    \label{eq: most general line element for spherically symmetric systems}
\end{equation}
where $d\Omega^2 = d\theta^2 + \sin^2(\theta)d\varphi^2$ is the metric on $S^2$. We show that it is possible to find a coordinate transformation
\begin{equation}
    r = E(R,T),\,\,\,\, t = F(R,T),
\end{equation}
so that $g_{rt} = 0$ and $D(r,t) = R^2$. Indicating with a subscript the partial differentiation with respect to one of the variables 
\begin{equation}
\begin{aligned}
        dr = E_R dR + E_T dT,\\
    dt = F_R dR + F_T dT,
\end{aligned}
\end{equation}
the line element (\ref{eq: most general line element for spherically symmetric systems}) in terms of the new coordinates reads
\begin{equation}
\begin{aligned}
    ds^2 = -& (B (E,F)F_T^2 -  A (E,F)E_T^2 - F_T E_T C(E,F) )dT^2 \\ 
            +& (A(E,F) E_R^2 - B(E,F)F_R^2 + F_R E_R C(E,F))dR^2  \\ +
            &(2 A(E,F)E_R E_T - 2 B(E,F)F_R F_T \\+& C(E,F)(F_R E_T + F_T E_R))dRdT + D(E,F)d\Omega^2.
\end{aligned}
\end{equation}
We look for a transformation such that 
\begin{equation}
\begin{cases}
\begin{aligned}
    2A(E,F)E_RE_T - 2B(E,F)F_RF_T +\\+ C(E,F)(F_RE_T + F_TE_R) = 0,
\end{aligned}\\
        D(E,F) = R^2.
\end{cases}
\label{eq: system of conditions for the transformation}
\end{equation}
Differentiating the second in $R$ and $T$ we get the conditions
\begin{equation}
    \begin{cases}
        E_R = 2R D_E^{-1} - F_R D_FD_E^{-1},\\
        E_T = -F_TD_FD_E^{-1}.
    \end{cases}
    \label{eq: conditions on E}
\end{equation}
Inserting in the first equation we get the conditions on $F_R,F_T$
\begin{equation}
\begin{cases}
    F_R = 2RD_E^{-1}(C-2AD_F)(2AD_F^2D_E^{-2}+2B+2CD_FD_E^{-1}),\\
    \forall F_T \neq 0.
\end{cases}
\label{eq: conditions on F}
\end{equation}
The system (\ref{eq: conditions on E}) can be integrated if 
\begin{equation}
    \begin{cases}
        E_{RT} = - F_{RT}D_FD_E^{-1},\\
        E_{TR} = - F_{TR}D_FD_E^{-1},
    \end{cases}
\end{equation}
which is satisfied if the integrability condition for (\ref{eq: conditions on F}) is met
\begin{equation}
    \begin{cases}
        F_{RT} = 0, \\
        F_{TR} = 0.
    \end{cases}
\end{equation}
We deduce that the system (\ref{eq: system of conditions for the transformation}) always has a solution with 
\begin{equation}
    F(R,T) = F_1(R) + F_2(T),
\end{equation}
for any choice of $F_2(T)$. This means that we can restrict to transformations of the type
\begin{equation}
    r = E(R,T),\,\,\, t = T + F_1(R),
\end{equation}
with the additional freedom in the time reparametrization 
$T \to T' = G(T)$.
This residual freedom was expected, since we require that the resulting time-space component of the metric vanishes. 
In conclusion, for a spherically symmetric system it is always possible to consider a coordinate system in which 
\begin{equation}
    ds^2 = -B(r,t)dt^2 + A(r,t)dr^2 + r^2 d\Omega^2,
    \label{eq2: eq: line element of a spherical system in proper coordinates -2}
\end{equation}
with $A,B > 0$. Equivalently, 
\begin{equation}
 ds^2 = -e^{\nu(r,t)}dt^2 + e^{\mu(r,t)}dr^2 + r^2 d\Omega^2.
\end{equation}
The expression of the Ricci scalar for this metric, is
\begin{equation}
\begin{aligned}
    R =& \frac{1}{2}e^{-\nu}(-\mu_t\nu_t +\mu_t^2 + 2\mu_{tt}) \\+& \frac{1}{2}e^{-\mu}\left(\mu_r\nu_r + \frac{4}{r}\mu_r- \nu_r^2 - \frac{4}{r}\nu_r - 2\nu_{rr}\right) +  \frac{2}{r^2} - \frac{2}{r^2}e^{-\mu}.
    \label{eq: Ricci scalar for the metric of a spherical system}
\end{aligned}
\end{equation}

\section{PRECESSION INTEGRAL}
\label{appendix: precession integral}
In this section we show in more detail the calculation of the integral (\ref{eq:integral for f(R) correction to the precession of the perihelion}). 

Introducing the variables $k \equiv \frac{a}{b}$ and $y \equiv \frac{a}{r}$, the integral can be restated as
\begin{multline}
    \varphi(b)|_{f(R)} = \frac{1}{4M_\odot G}\frac{c_1}{a^{n-1}}(1+k) \\\times \int_k^1\frac{dy}{(1-y)^{\frac{3}{2}} (y-k)^{\frac{3}{2}}}\left(\frac{k^n-k}{1-k} + \frac{1-k^n}{1-k}y - y^n\right).
    \label{eq: integrale for precession of perihelion after change of variables to y}
\end{multline}
Notice that the argument of the square brackets can be written as 
\begin{equation}
    [\dots] = (y-k)(1-y)\sum_{j = 0}^{n-2}k^j\sum_{l = 0}^{n-j-2}y^l,
\end{equation}
showing that the integral is convergent. Consider the integral
\begin{equation}
   I^{(m)} =  \int_k^1\frac{dy}{(1-y)^{a} (y-k)^{a}}y^m.
\end{equation}
The expression (\ref{eq: integrale for precession of perihelion after change of variables to y}) is the sum of three contributions of this type. The integral $I$ can be performed analytically for $a < 1$ and expressed in terms of a regularized, ordinary hypergeometric function 
\begin{equation}
    I^{(m)} = (1-k)^{1-2a}\frac{\Gamma(1-a)^2}{\Gamma(2-2a)}\ _2F_1(1-a,-m; 2-2a;1-k). 
\end{equation}
Expanding $_2F_1$ in terms of Euler $\Gamma$ functions $I$ can be written as
\begin{equation}
    I^{(m)} = (1-k)^{1-2a}\sum_{j=0}^m \binom{m}{j}\frac{\Gamma(1-a)\Gamma(1-a+j)}{\Gamma(2-2a+j)}(k-1)^j.
\end{equation}
Using this result, after some manipulations the combination 
\begin{equation}
    I = \frac{k^n-k}{1-k}I^{(0)} + \frac{1-k^n}{1-k}I^{(1)} - I^{(n)}
\end{equation}
can be expressed as
\begin{multline}
    I = (1-k)^{1-2a}\frac{\Gamma(1-a)^2}{\Gamma(2-2a)}\left[ 
    \vphantom{\sum_{j = 2}^n \binom{n}{j}\frac{\Gamma(2-2a)\Gamma(1-a+j)}{\Gamma(1-a)\Gamma(2-2a+j)}}
    -\frac{1}{2}(1-k^n) - \frac{1}{2}n(k-1)\right.\\\left. - \sum_{j = 2}^n \binom{n}{j}\frac{\Gamma(2-2a)\Gamma(1-a+j)}{\Gamma(1-a)\Gamma(2-2a+j)}(k-1)^j \right].
\end{multline}
The whole point is to take $\lim\limits_{a \to \frac{3}{2}} I$, since we know that the integral is well defined in this limit. Using that $\lim\limits_{z\to-1}\frac{1}{\Gamma(z)} =0$ we get 
\begin{equation}
    I = 2\sqrt{\pi}\sum_{j = 0}^{n-2}\binom{n}{j+2}\frac{\Gamma\left(j+\frac{3}{2}\right)}{\Gamma(j+1)}(k-1)^j
\end{equation}
and after some manipulations
\begin{multline}
    \varphi(b)|_{f(R)} = \frac{\pi}{4M_\odot G}\frac{c_1}{a^{n-1}}(1+k)\\ \times \sum_{j = 0}^{n-2}\binom{n}{j+2}\frac{(2j+1)!}{(j!)^2}\left(\frac{k-1}{4}\right)^j,
\end{multline}
which is the result reported in Eq.\ (\ref{eq: final formula for the leading correction introduced by f(R) to the angular precession of a closed orbit}).

\end{document}